\documentclass[twocolumn]{aastex62}

\newcommand{\myr}{$M_{\odot}$ yr$^{-1}$}
\newcommand{\msun}{$M_{\odot}$}
\newcommand{\mstar}{$M_{\star}$}

\newcommand{\sigmavel}{$\sigma_\star$}

\newcommand{\jwst}{\textit{JWST}}
\newcommand{\hst}{\textit{HST}}
\newcommand{\um}{$\mu$m}
\newcommand{\zphot}{$z_{\rm phot}$}

\usepackage{multirow}
\usepackage{booktabs}

\graphicspath{{./}{figures/}}

\submitjournal{ApJ}

\shorttitle{An atlas of quiescent galaxies in \jwst\ fields}
\shortauthors{Valentino et al.}

\begin{document}

\title{An Atlas of Color-selected Quiescent Galaxies at $z>3$ in Public \jwst\ Fields}

\correspondingauthor{Francesco Valentino}
\email{francesco.valentino@eso.org}

\author[0000-0001-6477-4011]{Francesco Valentino}
\affiliation{Cosmic Dawn Center (DAWN), Denmark}
\affiliation{Niels Bohr Institute, University of Copenhagen, Jagtvej 128, DK-2200 Copenhagen N, Denmark}
\affiliation{European Southern Observatory, Karl-Schwarzschild-Str. 2, D-85748 Garching bei Munchen, Germany}

\author[0000-0003-2680-005X]{Gabriel Brammer}
\affiliation{Cosmic Dawn Center (DAWN), Denmark}
\affiliation{Niels Bohr Institute, University of Copenhagen, Jagtvej 128, DK-2200 Copenhagen N, Denmark}

\author[0000-0003-4196-5960]{Katriona M. L. Gould}
\affiliation{Cosmic Dawn Center (DAWN), Denmark}
\affiliation{Niels Bohr Institute, University of Copenhagen, Jagtvej 128, DK-2200 Copenhagen N, Denmark}

\author[0000-0002-5588-9156]{Vasily Kokorev}
\affiliation{Kapteyn Astronomical Institute, University of Groningen, P.O. Box 800, 9700AV Groningen, The Netherlands}
\affiliation{Cosmic Dawn Center (DAWN), Denmark}

\author[0000-0001-7201-5066]{Seiji Fujimoto}
\affiliation{Department of Astronomy, The University of Texas at Austin, Austin, TX, USA}
\affiliation{Cosmic Dawn Center (DAWN), Denmark}

\author[0000-0002-8896-6496]{Christian Kragh Jespersen}
\affiliation{Department of Astrophysical Sciences, Princeton University, Princeton, NJ 08544, USA}

\author[0000-0002-1905-4194]{Aswin P. Vijayan}
\affiliation{Cosmic Dawn Center (DAWN), Denmark}
\affiliation{DTU-Space, Technical University of Denmark, Elektrovej 327, 2800, Kgs. Lyngby, Denmark}

\author[0000-0003-1614-196X]{John R. Weaver}
\affiliation{Department of Astronomy, University of Massachusetts, Amherst, MA 01003, USA}

\author[0000-0002-9453-0381]{Kei~Ito}
\affiliation{Department of Astronomy, School of Science, The University of Tokyo, 7-3-1, Hongo, Bunkyo-ku, Tokyo, 113-0033, Japan}

\author[0000-0002-5011-5178]{Masayuki Tanaka}
\affiliation{National Astronomical Observatory of Japan, 2-21-1 Osawa, Mitaka, Tokyo, 181-8588, Japan}
\affiliation{Department of Astronomical Science, The Graduate University for Advanced Studies, SOKENDAI, 2-21-1 Osawa, Mitaka, Tokyo, 181-8588, Japan}

\author[0000-0002-7303-4397]{Olivier Ilbert}
\affiliation{Aix Marseille Univ, CNRS, CNES, LAM, Marseille, France}

\author[0000-0002-4872-2294]{Georgios E. Magdis}
\affiliation{Cosmic Dawn Center (DAWN), Denmark}
\affiliation{DTU-Space, Technical University of Denmark, Elektrovej 327, 2800, Kgs. Lyngby, Denmark}
\affiliation{Niels Bohr Institute, University of Copenhagen, Jagtvej 128, DK-2200 Copenhagen N, Denmark}

\author[0000-0001-7160-3632]{Katherine E. Whitaker}
\affiliation{Department of Astronomy, University of Massachusetts, Amherst, MA 01003, USA}
\affiliation{Cosmic Dawn Center (DAWN), Denmark}

\author[0000-0002-9382-9832]{Andreas L. Faisst}
\affiliation{Caltech/IPAC, MS314-6, 1200 E. California Blvd. Pasadena, CA 91125, USA}

\author[0000-0002-9656-1800]{Anna Gallazzi}
\affiliation{INAF - Osservatorio Astrofisico di Arcetri, Largo Enrico Fermi 5, I-50125 Firenze, Italy}

\author[0000-0001-9885-4589]{Steven Gillman}
\affiliation{Cosmic Dawn Center (DAWN), Denmark}
\affiliation{DTU-Space, Technical University of Denmark, Elektrovej 327, 2800, Kgs. Lyngby, Denmark}

\author[0000-0001-9419-9505]{Clara Gim\'{e}nez-Arteaga}
\affiliation{Cosmic Dawn Center (DAWN), Denmark}
\affiliation{Niels Bohr Institute, University of Copenhagen, Jagtvej 128, DK-2200 Copenhagen N, Denmark}

\author[0000-0002-4085-9165]{Carlos G\'{o}mez-Guijarro}
\affiliation{Universit{\'e} Paris-Saclay, Universit{\'e} Paris Cit{\'e}, CEA, CNRS, AIM, 91191, Gif-sur-Yvette, France}

\author[0000-0002-7598-5292]{Mariko Kubo}
\affiliation{Astronomical Institute, Tohoku University, Aramaki, Aoba-ku, Sendai, Miyagi 980-8578, Japan}
\affiliation{Research Center for Space and Cosmic Evolution, Ehime University, 2-5 Bunkyo-cho, Matsuyama, Ehime 790-8577, Japan}

\author[0000-0002-9389-7413]{Kasper~E.~Heintz}
\affiliation{Cosmic Dawn Center (DAWN), Denmark}
\affiliation{Niels Bohr Institute, University of Copenhagen, Jagtvej 128, DK-2200 Copenhagen N, Denmark}

\author[0000-0002-3301-3321]{Michaela Hirschmann}
\affiliation{Institute for Physics, Laboratory for Galaxy Evolution and Spectral modelling, Ecole Polytechnique Federale de Lausanne, Observatoire de Sauverny, Chemin Pegasi 51, 1290 Versoix, Switzerland}
\affiliation{INAF - Osservatorio Astronomico di Trieste, Via Tiepolo 11, 34131 Trieste, Italy}

\author[0000-0001-5851-6649]{Pascal Oesch}
\affiliation{Department of Astronomy, University of Geneva, Chemin Pegasi 51, 1290 Versoix, Switzerland}
\affiliation{Cosmic Dawn Center (DAWN), Denmark}
\affiliation{Niels Bohr Institute, University of Copenhagen, Jagtvej 128, DK-2200 Copenhagen N, Denmark}

\author[0000-0003-3228-7264]{Masato Onodera}
\affiliation{Subaru Telescope, National Astronomical Observatory of Japan, National Institutes of Natural Sciences (NINS), 650 North A'ohoku Place, Hilo, HI 96720, USA}
\affiliation{Department of Astronomical Science, The Graduate University for Advanced Studies, SOKENDAI, 2-21-1 Osawa, Mitaka, Tokyo, 181-8588, Japan}

\author[0000-0001-9705-2461]{Francesca Rizzo}
\affiliation{Cosmic Dawn Center (DAWN), Denmark}
\affiliation{Niels Bohr Institute, University of Copenhagen, Jagtvej 128, DK-2200 Copenhagen N, Denmark}

\author[0000-0002-2419-3068]{Minju Lee}
\affiliation{Cosmic Dawn Center (DAWN), Denmark}
\affiliation{DTU-Space, Technical University of Denmark, Elektrovej 327, 2800, Kgs. Lyngby, Denmark}

\author[0000-0002-6338-7295]{Victoria Strait}
\affiliation{Cosmic Dawn Center (DAWN), Denmark}
\affiliation{Niels Bohr Institute, University of Copenhagen, Jagtvej 128, DK-2200 Copenhagen N, Denmark}

\author[0000-0003-3631-7176]{Sune Toft}
\affiliation{Cosmic Dawn Center (DAWN), Denmark}
\affiliation{Niels Bohr Institute, University of Copenhagen, Jagtvej 128, DK-2200 Copenhagen N, Denmark}

\begin{abstract}
We present the results of a systematic search for candidate quiescent galaxies in the distant Universe in eleven \textit{JWST} fields with publicly available observations collected during the first three months of operations and covering an effective sky area of $\sim145$ arcmin$^2$. We homogeneously reduce the new \textit{JWST} data and combine them with existing observations from the \textit{Hubble Space Telescope}. 
We select a robust sample of $\sim80$ candidate quiescent and quenching galaxies at $3 < z < 5$ using two methods: (1) based on their rest-frame $UVJ$ colors, and (2) a novel quantitative approach based on Gaussian Mixture Modeling of the $NUV-U$, $U-V$, and $V-J$ rest-frame color space, which is more sensitive to recently quenched objects. We measure comoving number densities of massive ($M_\star\geq 10^{10.6} M_\odot$) quiescent galaxies consistent with previous estimates relying on ground-based observations, after homogenizing the results in the literature with our mass and redshift intervals. However, we find significant field-to-field variations of the number densities up to a factor of $2-3$, highlighting the effect of cosmic variance and suggesting the presence of overdensities of red quiescent galaxies at $z>3$, as it could be expected for highly clustered massive systems. Importantly, \jwst\ enables the robust identification of quenching/quiescent galaxy candidates at lower masses and higher redshifts than before, challenging standard formation scenarios. All data products, including the literature compilation, are made publicly available.   
\end{abstract}


\section{Introduction} 
\label{sec:introduction}

Over the last few years, the existence of a population of quenched and quiescent galaxies (QGs) at redshifts $z \sim 3-4$ \citep[e.g.,][]{fontana_2009, straatman_2014, spitler_2014} has been finally corroborated by the long sought after spectroscopic confirmations \citep{glazebrook_2017,schreiber_2018b,schreiber_2018c,tanaka_2019,valentino_2020a,forrest_2020a,forrest_2020b, d'eugenio_2020, d'eugenio_2021, kubo_2021, nanayakkara_2022}. The combination of spectra and deep photometry have allowed for a first assessment of the physical properties of the newly-found early QGs. These properties include suppressed and minimal residual star formation rates (SFR), also supported with long-wavelength observations \citep{santini_2019, santini_2021, suzuki_2022}; emission from active galactic nuclei potentially pointing at a co-evolution with or feedback from their central supermassive black holes \citep{marsan_2015, marsan_2017, ito_2022, kubo_2022}; stellar velocity dispersions (\sigmavel) and dynamical masses \citep{tanaka_2019, saracco_2020} with possible implications on their initial mass function \citep{esdaile_2021, forrest_2022}; very compact physical sizes and approximately spheroidal shapes \citep{kubo_2018, lustig_2021}; and evidence that their large-scale environment may perhaps be overdense \citep{kalita_2021, kubo_2021, mcconachie_2022, ito_2023}.\\ 

Particular attention has been given to the reconstruction of the history (formation, quenching, and subsequent passive evolution) of distant QGs. A rapid and intense burst of star formation -- compatible with that of bright sub-millimeter galaxies with depletion timescales of $\tau \lesssim 100$ Myr -- is thought to drive the early mass assembly of the most massive and rarest systems \citep{forrest_2020b} as established for $z\sim2$ QGs \citep{cimatti_2008,toft_2014,akhshik_2022}. However, a more steady stellar mass assembly at paces typical of galaxies on the main sequence at $z>4$ might explain the existence of at least a fraction of the first QGs, likely less massive \citep{valentino_2020a}. In this case, the population of dust-obscured ``\textit{Hubble}-dark'' or ``optically-faint'' sub-millimeter detected sources could represent a good pool of candidate progenitors \citep{wang_2019, williams_2019, barrufet_2022, nelson_2022, perez-gonzalez_2022}. These results stem from various approaches and their inherent uncertainties, such as the modeling of star formation histories (SFHs) with different recipes -- parametric or not \citep[K. Gould et al. in preparation]{schreiber_2018c,ciesla_2016,carnall_2018,carnall_2019,leja_2019_sfh, iyer_2019}, matching comoving number densities of descendants and progenitors, also including ``duty cycles'' (i.e., there have to be at least as many star-forming predecessors as quiescent remnants accounting for the time window in which such progenitors are detectable, \citealt{toft_2014, valentino_2020a, manning_2022, long_2022}) or clustering analyses \citep{wang_2019}.\\ 

\begin{deluxetable*}{lrrccc}
    \tablecaption{Properties of the observed fields with \jwst/NIRCam observations.\label{tab:table_obs}}
    \tablehead{
    \colhead{Field}&
    \colhead{R.A.}&
    \colhead{Dec.}&
    \colhead{Area}&
    \colhead{NIRCam depths}&
    \colhead{\textit{HST}}\\ 
    \colhead{}&
    \colhead{\small [deg]}&
    \colhead{\small [deg]}&
    \colhead{\small [arcmin$^2$]}&
    \colhead{\small [mag]}&
    \colhead{}
    }
    \startdata
         CEERS             & $214.88598$ & $52.89500$  & $34.7$& 
            $28.5\,/\,28.8\,/\,28.8\,/\,28.8\,/\,28.3$& Yes \\
         Stephan's Quintet & $339.00057$ & $33.95996$  & $35.0$\tablenotemark{a}& 
            $27.5\,/\,27.6\,/\,28.0\,/\,28.1\,/\,27.7$& No \\
         PRIMER            & $34.37792$  & $-5.14717$  & $21.9$& 
            $27.5\,/\,27.7\,/\,27.9\,/\,27.9\,/\,27.4$&  Yes \\
         NEP               & $260.73773$ & $65.78167$  & $9.7$& 
            $28.5\,/\,28.6\,/\,28.9\,/\,28.9\,/\,28.3$& Yes \\    
         J1235             & $188.96741$ & $4.92465$   & $9.0$& 
            $28.4\,/\,29.1\,/\,29.3\,/\,29.3\,/\,28.4$& No \\    
         GLASS             & $3.50145$   & $-30.33612$ & $8.5$& 
            $28.8\,/\,29.0\,/\,29.1\,/\,29.1\,/\,29.4$& Yes \\    
         Sunrise           & $24.34743$  & $-8.43215$ & $7.3$\tablenotemark{b}& 
            $28.1\,/\,28.3\,/\,28.4\,/\,28.4\,/\,28.0$& Yes \\    
         SMACS0723         & $110.75478$ & $-73.46788$& $6.5$\tablenotemark{b}& 
            $28.8\,/\,29.0\,/\,29.2\,/\,29.2\,/\,28.8$& Yes \\    
         SGAS1723          & $260.91450$ & $34.19371$ & $5.3$& 
            $25.8\,/\,25.9\,/\,26.6\,/\,26.7\,/\,26.6$& Yes \\            
         SPT0418           & $64.66113$ & $-47.87526$ & $5.0$& 
            $26.6\,/\,27.1\,/\,27.8\,/\,27.3\,/\,27.1$& No \\            
         SPT2147\tablenotemark{c} & $326.82917$ & $-50.59632$ & $2.3$& 
            --- $/\,27.7\,/\,27.4\,/\,27.7\,/\,26.9$& Yes \\                     
    \enddata
    \tablecomments{\textbf{NIRCam depths:} expressed as $5\sigma$ within the $0\farcs5$ apertures used for the photometric extraction in the area covered by F150W / F200W / F277W / F356W / F444W (Appendix \ref{appendix:fields}).}
    \tablenotetext{a}{The area covered by the group members has been masked (Appendix \ref{appendix:fields}).}
    \tablenotetext{b}{Effective area accounting for the gravitational lensing effect at $z \sim 3-5 $ (Section \ref{sec:number_densities}).}
    \tablenotetext{c}{No F150W coverage.}
\end{deluxetable*}

Debate continues on the exact mechanisms causing the cessation of the star formation at $z\gtrsim3-4$, as well as at other redshifts (\citealt{man_2018} for a compendium). However, at high redshift there is the significant advantage of observing such a young Universe that classical ``slow'' quenching processes operating on $\geq 1-2$ Gyr timescales at low redshifts are disfavored (e.g., strangulation or gas exhaustion \citealt{schawinski_2014, peng_2015}). Moreover, aided by sample selections favoring high detection rates over completeness, the best characterized spectroscopically confirmed QGs tend to show signatures of recent quenching ($\sim$ a few hundred Myr) as in ``post-starburst'' galaxies rather than being prototypical ``red and dead'' objects \citep{schreiber_2018c, d'eugenio_2020,forrest_2020b, lustig_2021, marsan_2022, gould_2022}, even if examples of older populations are available \citep[M. Tanaka et al. in preparation]{glazebrook_2017}. The analysis of larger samples of galaxies during or right after quenching could eventually help us understand the physics behind this phenomenon in the early Universe.\\ 

The exploration of post-quenching evolution is also in its infancy. There are indications of a simultaneous passive aging of the stellar populations and a rapid size evolution, but only modest stellar mass increase via dry minor mergers \citep{tanaka_2019}, resembling the second act of the popular ``two-phase'' evolutionary scenario that explains how $z\sim2$ QGs change over time \citep[e.g.,][]{delucia_2006, cimatti_2008, naab_2009, oser_2010}. From the point of view of stellar dynamics, the small sample of QGs with available velocity dispersions does not allow for drawing any strong conclusions about possible evolutionary paths at constant or time varying \sigmavel\ yet \citep{tanaka_2019, saracco_2020, esdaile_2021, forrest_2022}. \\

These first results already paint a rich picture of how the earliest QGs formed and quenched and indicate several promising research avenues to explore. However, they relied on the availability of deep near-infrared photometry and ground-based spectroscopy, which come with obvious limitations on the spatial resolution and wavelength coverage. So far, these prevented us from unambiguously confirming if QGs exist at $z>4$ (see \citealt{merlin_2019, carnall_2020, mawatari_2020} for possible candidates), clearly defining the first epoch of sustained galaxy quenching, and ascertaining the existence of low-mass systems potentially quenched by different processes.\\

\jwst\ enables us to break this ceiling, looking farther and deeper to catch the earliest QGs spanning a vast range of stellar masses. The first months of observations kept this promise and already offer a spectacular novel view on early galaxy evolution in general. In this work, we aim to capitalize on publicly available \jwst\ multi-wavelength imaging in 11 fields to find and quantify the population of early QGs, pushing the limits in redshift and mass affecting ground-based surveys. This paper is the first of a series addressing several of the contentious scientific points mentioned above. Here we will focus on the \jwst-based selection of a robust sample of photometric QG candidates and on the bare-bones comoving number density calculations, taking advantage of the coverage of a relatively large combined area of $\sim 145$ arcmin$^2$ at $z\sim3-5$ and the scattered distribution on the sky to reduce the impact of cosmic variance. Counting galaxies is a basic test for models and simulations and, in the case of distant QGs, it has generated quite some discussion on the robustness of current theoretical recipes \citep[e.g.,][]{schreiber_2018c, merlin_2019}. Also, accurate number densities are key ingredients to try to establish an evolutionary connection among populations across redshifts, thus affecting our view of the history of assembly of the first QGs.\\

The data collection, homogeneous reduction, and modeling are presented in Section \ref{sec:data}. Our \jwst-based color selection is described in Section \ref{sec:sampleselection}, followed by the results on number densities contextualized within the current research landscape in Section \ref{sec:number_densities}. Throughout the paper, we assume a $\Lambda$CDM cosmology with $\Omega_{\rm m} = 0.3$, $\Omega_{\Lambda} = 0.7$, and $H_0 = 70\,\mathrm{km\,s^{-1}\,Mpc^{-1}}$. All magnitudes are expressed in the AB system. All the reduced data, selected samples, and physical properties discussed in this work are publicly available online \footnote{Supplementary material and catalogs of selected sources: \dataset[10.5281/zenodo.7614908]{https://doi.org/10.5281/zenodo.7614908};\\ mosaics and field catalogs \dataset[10.17894/ucph.e3d897af-233a-4f01-a893-7b0fad1f66c2]{https://doi.org/10.17894/ucph.e3d897af-233a-4f01-a893-7b0fad1f66c2}\label{footnote:v4_archive_url}}.
\begin{figure*}
    \includegraphics[width=\textwidth]{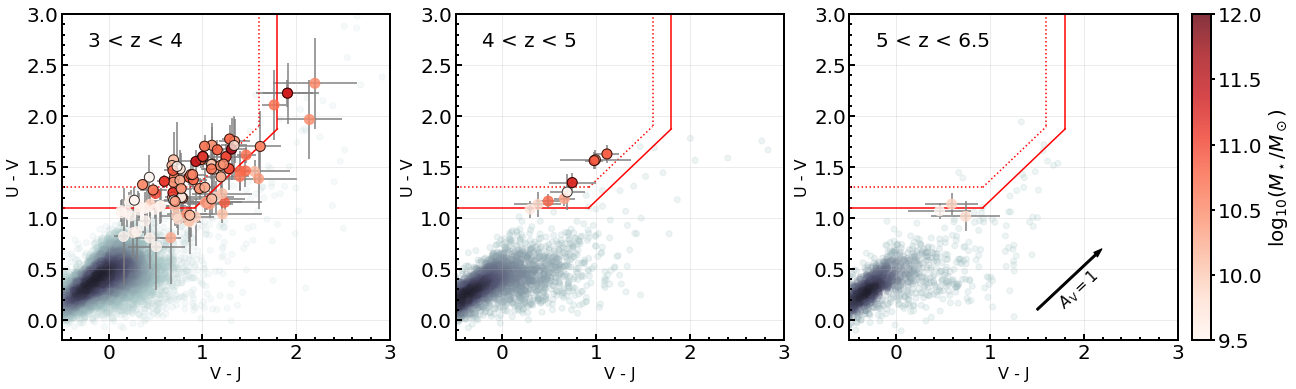}
    \caption{$U-V$, $V-J$ rest-frame color diagrams for our combined sample of galaxies in \jwst\ fields binned in redshift as labeled. Filled circles indicate our visually-inspected $UVJ$ quiescent sample and their $1\sigma$ uncertainties, color coded according to their stellar mass. We circled in black the object in the ``strict'' sample. Gray points indicate the rest of the sample at those redshifts (Section \ref{sec:sampleselection}). The color intensity scales as the density of points. The red dotted and solid lines indicate the standard selection box \citep{williams_2009} and a looser version allowing for an extra pad of 0.2 mag, respectively. The black arrow shows the effect of reddening for $A_{V}=1$.}
    \label{fig:uvj}
\end{figure*}

\section{Data}
\label{sec:data}
In the following sections, we present our reduction and analysis of the photometric data. A dedicated paper will describe all the details of this process (G. Brammer et al., in preparation). The approach is similar to that in \cite{labbe_2022} and \cite{bradley_2022}, here including the recently updated zeropoints.

\subsection{Reduction}
\label{subsec:reduction}
We homogeneously process the publicly available \jwst\ imaging obtained with the NIRCam, NIRISS, and MIRI instruments in $11$ fields targeted during the first three months of operations (Table \ref{tab:table_obs}). We retrieved the level-2 products and processed them with the \textsc{Grizli} pipeline \citep{brammer_2021, brammer_2022}. Particular care is given to the correction of NIRCam photometric zeropoints relative to \texttt{jwst\_0942.pmap}, including detector variations\footnote{\dataset[10.5281/zenodo.7143382]{https://doi.org/10.5281/zenodo.7143382}}. The results are consistent with similar efforts by other groups \citep{boyer_2022, nardiello_2022} and with the more recent \texttt{jwst\_0989.pmap} calibration data. Corrections and masking to reduce the effect of cosmic rays and stray light are also implemented \citep[see][]{bradley_2022}. For the PRIMER data, we introduce an additional procedure that alleviates the detrimental effects of the diagonal striping seen in some exposures. Finally, our mosaics include the updated sky flats for all NIRCam filters.  We further incorporate the available optical and near-infrared data available in the Complete \textit{Hubble} Archive for Galaxy Evolution \citep[CHArGE,][]{kokorev_2022}. We align the images to Gaia DR3 \citep{gaia-collaboration_2021}, co-add, and finally drizzle them \citep{fruchter_2002} to a $0\farcs02$ pixel scale for the Short Wavelength (SW) NIRCam bands and to $0\farcs04$ for all the remaining \jwst\ and \textit{Hubble Space Telescope} (\hst) filters. We provide further details about the individual fields in Appendix \ref{appendix:fields}.\\

\subsection{Extraction}
\label{subsec:extraction}
We extract sources using a detection image produced by combining of all the ``wide'' (W) NIRCam Long Wavelength (LW) filters available (typically F277W+F356W+F444W) optimally weighted by their noise maps. For source extraction, we use \textsc{sep} \citep{barbary_2016}, a pythonic version of \textsc{source extractor} \citep{bertin_1996}. We extract the photometry in circular apertures with a diameter of $0\farcs5$ and correct to the ``total'' values within an elliptical Kron aperture \citep{kron_1980}\footnote{Photometric measurements in different apertures are available in the online catalogs\textsuperscript{\ref{footnote:v4_archive_url}}.}. 
The aperture correction is computed on the LW detection image and applied to all bands. The depth in the reference $0\farcs5$ apertures in the five NIRCam bands that we require to select candidate quiescent galaxies (F150W, F200W, F277W, F356W, and F444W, Section \ref{sec:sampleselection}) are reported in Table \ref{tab:table_obs}. The galaxy distribution as a function of redshift for F444W is shown in Appendix (Figure \ref{fig:f444w_limits}) and for the remaining bands in Figure Set A1. An extra-correction of $\sim10$\% to account for the flux outside the Kron aperture in \hst\ bands and optimal for point-like sources is computed by analyzing the curve of growth of point spread functions (PSF)\footnote{\url{https://www.stsci.edu/hst/instrumentation/wfc3/data-analysis/psf}}.

\subsection{Modeling of the spectral energy distribution}
\label{subsec:modeling}
We utilize \textsc{eazy-py}\footnote{\url{https://github.com/gbrammer/eazy-py}} \citep{brammer_2008} to estimate photometric redshifts, rest-frame colors, and stellar masses from the $0\farcs5$ diameter aperture photometry corrected to total fluxes as described above. We apply residual zeropoint corrections to optimize the photometric redshifts with solutions free to vary in the interval $z=0-18$. We use the same set of 13 templates from the Flexible Stellar Populations Synthesis code \citep[\textsc{fsps},][]{conroy_2010} described in \cite{kokorev_2022} and \citet{gould_2022}, linearly combined to allow for the maximum flexibility. This set of templates covers a large interval in ages, dust attenuation, and log-normal star formation histories -- spanning the whole $UVJ$ rest-frame color diagram. More specifically, the \texttt{corr\_sfhz\_13} subset of models within \textsc{eazy} contains redshift-dependent SFHs, which, at a given redshift, exclude histories that start earlier than the age of the Universe. A template derived from the NIRSpec spectrum of a confirmed strong line emitter at $z=8.5$ -- ID4590 from \cite{carnall_2022a} -- is also included to allow for an extra degree of freedom in photometric solutions of distant objects, but not accounted for in the stellar mass calculation\footnote{The parameters attached to this template are uncertain \citep{carnall_2022a,gimenez-arteaga_2022}. Solutions dominated by this template should be treated with caution. This is not the case of galaxies in our final samples.}. The templates are created adopting a \citet{chabrier_2003} initial mass function and applying a \citet{kriek_2013} dust attenuation law (dust index $\delta=-0.1$, $R_{\rm V}=3.1$), where the maximum allowed attenuation is also redshift-dependent. 
A fixed grid of nebular emission lines and continuum from \textsc{cloudy} (v13.03) models is added to the templates within \textsc{fsps} (metallicity: $\mathrm{log}(Z/Z_\odot)\in [-1.2,0]$, ionization parameter $\mathrm{log}(U)=-1.64,-2$; \citealt{byler_2018_cloudyfsps}). Given their fixed ratios and sole purpose of modeling the photometry, the strength of the emission lines should be taken with caution. We, thus, do not include them in our analysis. The templates, their input parameters, and the redshift evolution of their allowed SFHs and attenuation are available online\footnote{\url{https://github.com/gbrammer/eazy-photoz/tree/master/templates/sfhz}}. Also, a correction for the effect of dust in the Milky Way is applied to the templates within \textsc{eazy-py}, pulling the Galactic dust map by \cite{schlafly_2011} from \textsc{dustmaps} \citep{green_2018}. This effect is relevant for SMACS0723 \citep[$E(B-V)_{\rm MW}=0.19$, see also][]{faisst_2022} and to a lesser extent for the rest of the fields ($E(B-V)_{\rm MW}=0.007-0.07$). In terms of photometric redshifts, we obtain a good agreement with spectroscopic determinations from archival observations when available (G. Brammer et al., in preparation). For reference, in the CEERS field we estimate a $\sigma_{\rm NMAD}=0.0268\,(0.0187)$ for the spectroscopic sample (clipping catastrophic outliers), respectively\footnote{The redshift distributions and the comparison with spectroscopic estimates (e.g., $\Delta z/(1+z)$ as a function of $z$) for each field are bundled with the mosaics\textsuperscript{\ref{footnote:v4_archive_url}}.}. Stellar masses are consistent with independent estimates obtained with finer grid-based codes (Figure \ref{fig:3dhst} in Appendix for a comparison with 3D-$HST$). Star formation rates are also found in agreement with determinations with alternative codes at $z=0.5-3$ \citep{gould_2022}. However, we opt not to rely on SFR for our selection and analysis at $z>3$ to adhere as close as possible to observables.
\begin{figure*}
    \includegraphics[width=\textwidth]{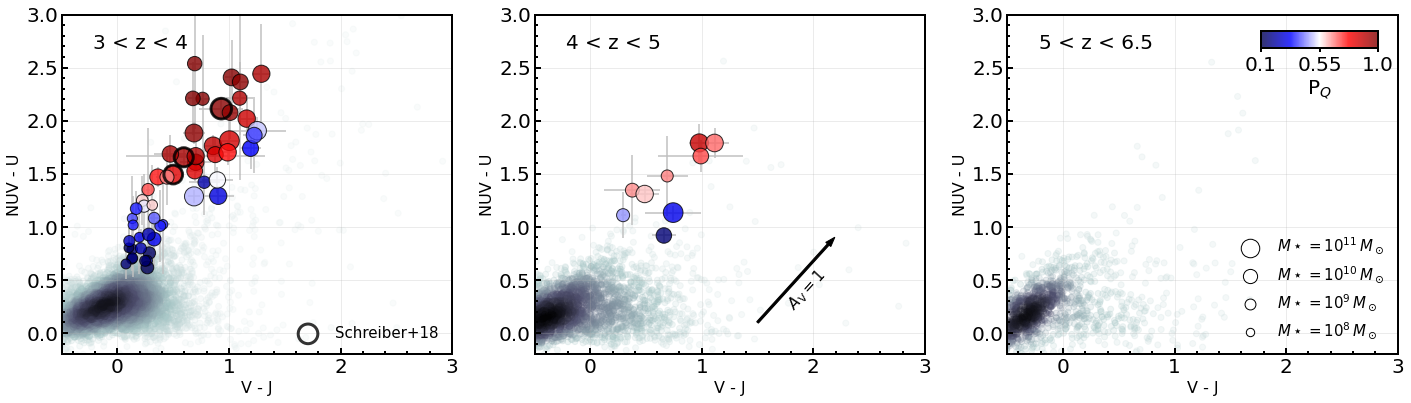}
    \caption{$NUV-U$, $V-J$ rest-frame color diagrams for our combined sample of galaxies in \jwst\ fields binned in redshift as labeled. Filled circles indicate our robust $NUVUVJ$-selected quiescent sample ($P_{\rm Q, 50\%}\geq0.1$) and their $1\sigma$ uncertainties, color coded according to the nominal probability of being quiescent $P_{\rm Q}$ for display purposes. The symbol size scales proportionally to the stellar mass as labeled. Thicker black circles show the sources with a robust or uncertain $z_{\rm spec}$ in \cite{schreiber_2018c} falling in the portion of the CEERS field considered here. Gray points indicate the rest of the sample at those redshifts (Section \ref{sec:sampleselection}). The color intensity scales as the density of points. The black arrow shows the effect of reddening for $A_{V}=1$.}
    \label{fig:nuvuvj}
\end{figure*}

\subsection{Rest-frame colors}
\label{subsec:rfcolors}
As described in \cite{gould_2022}, besides photometric redshifts constrained by spectral features, \textsc{eazy-py} returns the physical quantities attached to each template, propagated through the minimization process and computed using the same set of coefficients providing the best-fit $z_{\rm phot}$. Uncertainties are estimated at $z_{\rm phot}$ as the 16-84\% percentiles of a 100 fits drawn from the best-fit template error function. We compute the rest-frame magnitudes in the GALEX $NUV$ band ($\lambda = 2800$ \AA), the $U$ and $V$ Johnson filters defined as in \citet{maiz-apellaniz_2006}, and the $J$ 2MASS passband \citep{skrutskie_2006}. The rest-frame magnitudes are computed following a hybrid approach that uses the templates as guides for a weighted interpolation of the observations and accounts for bandpasses and relative depths (\citealt{brammer_2008, brammer_2011}, and Appendix A in \citealt{gould_2022}). This allows for a color determination which relieves the dependency on the adopted models, while using the whole photometric information.

\section{Sample selection}
\label{sec:sampleselection}
Before selecting quiescent galaxy candidates at $z>3$, we start by applying a series of loose cuts to immediately reject galaxies with unreliable photometric modeling. We constrain the quality of the fit to $\chi^2/N_{\rm filt}\leq 8$, where $N_{\rm filt}\geq 6$ is the number of available filters. The latter includes NIRCam wide bands at $1.5,\,2.0,\,2.7,\,3.5,\,4.4$ \um\ in every field with the exception of SPT2147, where imaging with F150W was not taken. Coupled with the adoption of the NIRCam LW detection image (Section \ref{subsec:reduction}), this requirement enforces a selection based on \jwst\ data, while the coverage at observed wavelengths longer than $3$ \um\ allows for robust determinations of stellar masses. Then we apply a loose cut at $>5$ on the quadratic sum of the S/N of the aperture fluxes in these NIRCam bands. We constrain the location of the peak and the tightness of the redshift probability distribution function $p(z)$ ($\mathrm{max}\{p(z)\}>0.5$, $(p(z)_{84\%}-p(z)_{16\%})/(2\,p(z)_{50\%})<0.3$, where $p(z)_{i\%}$ indicate the $i$-th percentile of $p(z)$). Finally, we apply a cut in redshift at $3 \leq z \leq 6.5$ with a buffer of $dz=0.1$ and accounting for the uncertainty on the best-fit solution ($p(z)_{84\%}\geq3-dz$ and $p(z)_{16\%}\leq 6.5+dz$). To pick quiescent objects, we opt for a rest-frame color selection following a dual approach. 

\subsection{$UVJ$ color diagram}
\label{subsec:uvj}
On the one hand, we select objects in the classical $UVJ$ diagram. This allows us to directly compare our results with a large body of literature that has accumulated over the last two decades. We allow for a 0.2 mag extra pad when compared with the cuts in \cite{williams_2009} and we initially retain sources with $1\sigma$ uncertainties on the colors consistent with the selection box as long as $\sigma_{\rm color}<0.5$ mag. We then visually inspect the images and the SED fits of $251$ candidate quiescent galaxies. We retain $109/251$ objects ($\sim45$\%) after excluding remaining bad fits or poor quality images affected by edge effects, spikes, or contaminating bright sources. We show the location of the visually inspected sample in three redshift bins at $z>3$ in Figure \ref{fig:uvj}. The visual selection significantly shrinks the initial pool of candidates. This is expected given the deliberate choice of starting from rather loose constraints not to lose potential good candidates. The visual cut particularly hits the highest redshift pool of candidates: we retain $3/56$ galaxies at $z>5$ largely due to poor quality SEDs. For transparency, all of the SEDs and cutouts of the discarded sources during the visual check are also released.\\  

To draw straightforward comparisons with previous works and in attempt to remedy the larger contamination that inevitably affects our expanded selection box, we further flag our sources as ``strict'' and ``padded''. The first tag refers to $55/109$ sources that fall in the classical QG box, also accounting for their $1\sigma$ uncertainties ($34/109$ without including the latter as in the ``standard'' selection). The second flag refers to $67/109$ sources that have nominal (i.e., without including uncertainties) colors within the 0.2 mag padded locus of QGs. The overlapping sample comprises $51$ galaxies. Differences in the derived number densities primarily reflect these further distinctions.\\

Three-color NIRCam SW and LW cutouts, photometry, SED models, and basic properties estimated with \textsc{eazy-py} are publicly available for the $UVJ$-selected samples.  

\subsection{$NUV-U, V-J$ color diagram}
\label{subsec:nuvuvj}
In parallel, we follow the novel method described in \citet{gould_2022} (see also \citealt{antwi-danso_2022} for an alternative approach introducing a synthetic band). The authors incorporate the $NUV$ magnitude in their selection and model the galaxy distribution in the ($NUV-U$, $U-V$, $V-J$) space with a minimal number of Gaussians carrying information (Gaussian Mixture Model, GMM, \citealt{pedregosa_2011}). The addition of the $NUV$ magnitude makes the selection more sensitive to recent star formation and, thus, to recently quenched or post-starburst objects \citep{arnouts_2013, leja_2019}, which are expected to be observed at high redshift as we approach the epoch of quenching of the first galaxies \citep{d'eugenio_2020, forrest_2020b}. Moreover, the GMM allows to fully account for the blurred separation between star-forming and quiescent galaxies at $z>3$, assigning a ``probability of being quiescent'' $P_{\rm Q}$ to each object and bypassing the use of arbitrary color cuts. The GMM grid is calibrated on a sample of $2<z<3$ galaxies in the COSMOS2020 catalog \citep{weaver_2022} assuming $5\times$ more conservative \textit{Spitzer}/IRAC uncertainties and refit with \textsc{eazy-py} in a similar configuration to that adopted here \citep{valentino_2022}. To account for the uncertainties on the colors, we bootstrap their values 1000 times and use the median and 16-84\% percentiles of the distribution as our reference $P_{\rm Q, 50\%}$\footnote{For clarity, we stress that, in this notation, ``50\%'' refers to the percentile of the boostrapped $P_{\rm Q}$ and not to a probability of 50\% to be quiescent.} and its $1\sigma$ uncertainties. We also list the nominal $P_{\rm Q}$ associated with the best-fit colors in our catalogs.\\

In the rest of our analysis, we adopt a cut at $P_{\rm Q, 50\%}\geq0.1$ to select candidate quiescent galaxies, with a threshold set at $P_{\rm Q, 50\%}=0.7$ to separate passive galaxies from objects showing features compatible with more recent quenching (see \citealt{gould_2022} for a description of the performances of different cuts benchmarked against simulations). As for the $UVJ$-selected sample, we visually inspect all of the images and SEDs of the candidates that made our initial $P_{\rm Q, 50\%}$ cut. Finally, we retain $50/71$ sources ($70$\%) with $0.1 \leq P_{\rm Q, 50\%}<0.7$ and $18/20$ ($\sim90$\%) truly passive galaxy candidates with $P_{\rm Q, 50\%}\geq0.7$. Their location in the projected $NUV-U$, $V-J$ plane is shown in Figure \ref{fig:nuvuvj}. 

\subsection{Overlap between selections}
\label{subsec:overlap}
As noted in \citet{gould_2022}, a selection in the $NUVUVJ$ arguably outperforms the classical $UVJ$ in selecting quiescent (passive and recently quenched or post-starbust) galaxies at $z>3$. However, the two criteria partially overlap and identify the same quiescent sources -- to an extent fixed by the $P_{Q}$ threshold and the exact location of the selection box in the $UVJ$ diagram. The boundaries adopted here slightly differ from those in \citet{gould_2022}, but the resulting overlap between the selection criteria at $3<z<5$ is similar. Focusing on the visually inspected samples, $52$ sources are selected by both techniques. These amount to $\sim75$\% and $\sim50$\% of the $NUVUVJ$ and $UVJ$-selected objects, respectively, comparable with the fractions reported in \cite{gould_2022} in the same redshift range. In more detail, $100$\% and $\sim70$\% of the sources with $P_{\rm Q, 50\%}\geq0.7$ and $0.1 \leq P_{\rm Q, 50\%}<0.7$ are part of the $UVJ$ sample. Moreover, $16/18$ objects with $P_{\rm Q, 50\%}\geq0.7$ fall within the standard $UVJ$ selection box. This is because the galaxies assigned lower $P_{\rm Q, 50\%}$ values are those in the region bordering star-forming and quiescent, whereas galaxies with higher $P_{\rm Q, 50\%}$ are those which resemble more classically quiescent galaxies owing to how the model is trained (Figure \ref{fig:nuvuvj}). Therefore, it is expected that the $UVJ$ selected sample has a smaller overlap with galaxies that have lower $P_{\rm Q, 50\%}$ values and the galaxies that it does not select are those which are recently quenched.
The overlap is reflected also in the \mstar\ distributions of the selected samples (Figure \ref{fig:mass_distribution} in Appendix). Lower $P_{\rm Q, 50\%}$ values are associated to bluer, more recently quenched, but also lower mass objects, otherwise missed by $UVJ$ selections.

\subsection{Sanity checks on the sample}
\label{subsec:sanity}
We test our selection and draw comparisons with what has been achieved before the advent of \jwst\ in a variety of ways, as summarized below. More details can be found in Appendix \ref{appendix:checks}. 

\subsubsection{Comparison with \hst-based photometry}
First, we compare our \hst/F160W photometry (consistent with that of NIRCam/F150W), photometric redshift, and stellar mass estimates against those from the 3D-\textit{HST} catalog \citep{skelton_2014} overlapping with part of the CEERS (EGS) and PRIMER (UDS) areas (Figure \ref{fig:3dhst}). Despite different detection images, for those sources in common we find excellent agreement in terms of aperture photometry and redshifts derived. Moreover, minimal systematic offsets in $\mathrm{log}(M_\star/M_\odot)$ ($<0.2$ dex) make our results fully consistent between different SED modeling codes (Appendix \ref{subappendix:3dhst}). 

\subsubsection{Availability of \hst\ imaging}
We also test the impact on our sample selection of \hst\ filters, which increase the sampling of the rest-frame UV/optical wavelengths in some of the fields (Table \ref{tab:coverage} in Appendix \ref{appendix:fields}). We refit the photometry in the CEERS and PRIMER fields retaining only the available NIRCam filters among those at 0.9, 1.15, 1.5, 2.0, 2.7, 3.5, and 4.4 $\mu$m, mimicking the situation for Stephan's Quintet where no \hst\ imaging is at disposal. We obtain fully consistent $z_{\rm phot}$, \mstar, and rest-frame color estimates within the uncertainties, especially when removing the effect of $z_{\rm phot}$ from the calculation and focusing on the $3 \leq z \leq6.5$ interval of interest (Figure \ref{fig:filters}). This holds also when F090W, probing wavelengths shorter than $NUV$ at the lower end of the redshift interval that we explore, is not available, as in the case of CEERS. We also re-apply the $NUVUVJ$ selection, including bootstrapping, and obtain consistent samples taking into account the uncertainties. 

\subsubsection{Dusty star-forming or high-redshift contaminants}
When available, we look for counterparts at long wavelengths to exclude obvious dusty interlopers. We search for matches in sub-millimetric observations in CEERS ($450$ and $850$ \um\ with Scuba-2, \citealt{zavala_2017, geach_2017}), PRIMER ($870$ \um\ with ALMA from the AS2UDS survey, \citealt{dudzeviciute_2020}, and \citealt{cheng_2022_pearls}), and SMACS0723 ($1.1$ mm with ALMA from the ALCS Survey, \citealt{kokorev_2022}; S. Fujimoto et al. in preparation). We found only one potential association with a $\sim5\sigma$ Scuba-2 detection at $850$ \um\ in CEERS (S2CLS-EGS-850.063 in \citealt{zavala_2017}, \#9329 in our catalog). This is a weak sub-millimeter galaxy candidate possibly associated with an overdensity (S. Gillman et al. in preparation). Removing it from our sample does not change the results of this work. The absence of long-wavelength emission in our pool of visually inspected candidates supports their robustness. We also note that the contamination of high-redshift Lyman Break Galaxies is negligible, given their number densities \citep{fujimoto_2022_dropouts}.

\subsubsection{Spectroscopically confirmed and alternative \jwst-based photometric samples}
We correctly identify and select the spectroscopically confirmed QGs in \cite{schreiber_2018c} at consistent $z_{\rm phot}$. Our relaxed $UVJ$ cuts and $P_{\rm Q,50\%}\geq0.1$ recover $14/17$ and $11/17$ candidate QGs selected via SED modeling and a sSFR threshold in \cite{carnall_2022qg}, respectively. When considering only their ``robust'' sample, there is a 89\% and 78\% overlap with our $UVJ$ and $P_{\rm Q,50\%}$ selections. We calculate lower, but generally consistent $z_{\rm phot}$ (Figure \ref{fig:carnall}, see also \citealt{kocevski_2022}). Minor systematic offsets in the \mstar\ and F200W magnitudes are present (Figure \ref{fig:carnall}).

\begin{figure*}
    \centering    
    \includegraphics[width=\textwidth,height=\textheight,keepaspectratio]{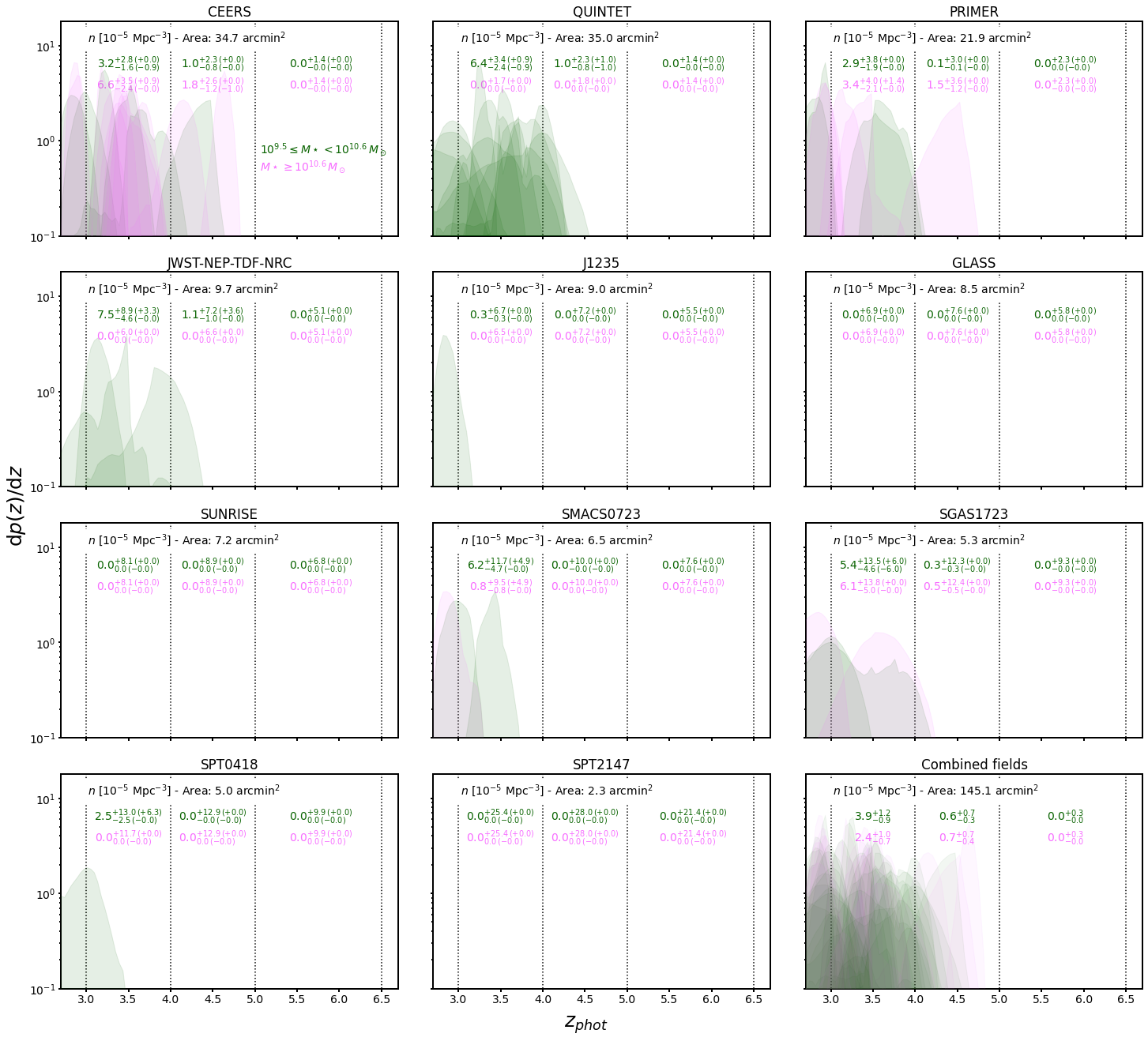}
    \caption{Number densities of $UVJ$-selected galaxies in public {\it JWST} fields. The purple and green colors mark the $p(z)$ of individual robust ``strict'' $UVJ$ quiescent candidates with $M_\star \geq 10^{10.6}\,M_\odot$ and $10^{9.5}\leq M_\star < 10^{10.6}\,M_\odot$, respectively. The $p(z)$ are normalized by their area ($\int_{z} p(z)\, \mathrm{d}z=1$).  The sky coverage of each field is as labeled. We included the effect of gravitational lensing at these redshifts in the calculation of the areas around galaxy clusters (Sunrise, SMACS0723), for which the masses should be intended as observed (not delensed). The comoving number densities in units of $10^{-5}$ Mpc$^{-3}$ obtained from the integration of $p(z)$ within the $3<z<4$, $4<z<5$, and $5<z<6.5$ bins marked by dotted lines are reported. The errors mark the 68\% Poissonian confidence intervals. Estimates of the uncertainties from bootstrapping are within brackets. The first and second rows indicate $n$ in the $10^{9.5}\leq M_\star < 10^{10.6}\,M_\odot$ and $M_\star \geq 10^{10.6}\,M_\odot$ bins, respectively.}
    \label{fig:uvj_number_densities}
\end{figure*}

\section{Number densities}
\label{sec:number_densities}
For each field, we compute the comoving number density $n$ of candidate quiescent galaxies in three redshift bins ($z\in [3,4)$, $[4,5)$, and $[5,6.5)$). We compute the number of objects per bin by integrating the $p(z)$, thus accounting for the uncertainties on the photometric redshift determination. As an alternative estimate of the statistical errors associated with the latter, we randomly extract $1000\times$ each $z_{\rm phot}$ within their $p(z)$ and compute the median and $16-84$\% percentiles, finding consistent results. We also compute the $68$\% Poissonian confidence intervals or upper limits \citep{gehrels_1986}. The comoving volumes are calculated starting from the area subtended by the observations and satisfying the requirements in terms of band coverage and minimum number of filters (Section \ref{sec:sampleselection}). 
For the Sunrise and SMACS0723 fields, we account for the effect of gravitational lensing on the volume at $z=3-5$ as in \cite{fujimoto_2016}. We estimate the intrinsic survey volume by producing magnification maps at $z=3,4$, and $5$. We base the calculation on the mass model constructed with the updated version of \textsc{glafic} \citep{oguri_2010, oguri_2021b} using the available \hst\ and \jwst\ data  \citep{harikane_2022, welch_2022c}. The effect of lensing on the effective area varies negligibly in the redshift interval and luminosity regime spanned by our samples of candidate quiescent galaxies. The linear magnification for the only QG candidate in proximity of SMACS0723 is $\sim 1.7$. For two candidates in WHL0137 (Sunrise) with $\mathrm{log}(M_\star/M_\odot)<9.5$ and $P_{\rm Q,50\%}<0.7$, this factor is $\sim3$. We did not apply the magnification correction to the parameter estimates. This does not affect the conclusions on number densities.\\

\begin{figure*}
    \centering    
    \includegraphics[width=\textwidth,height=\textheight,keepaspectratio]{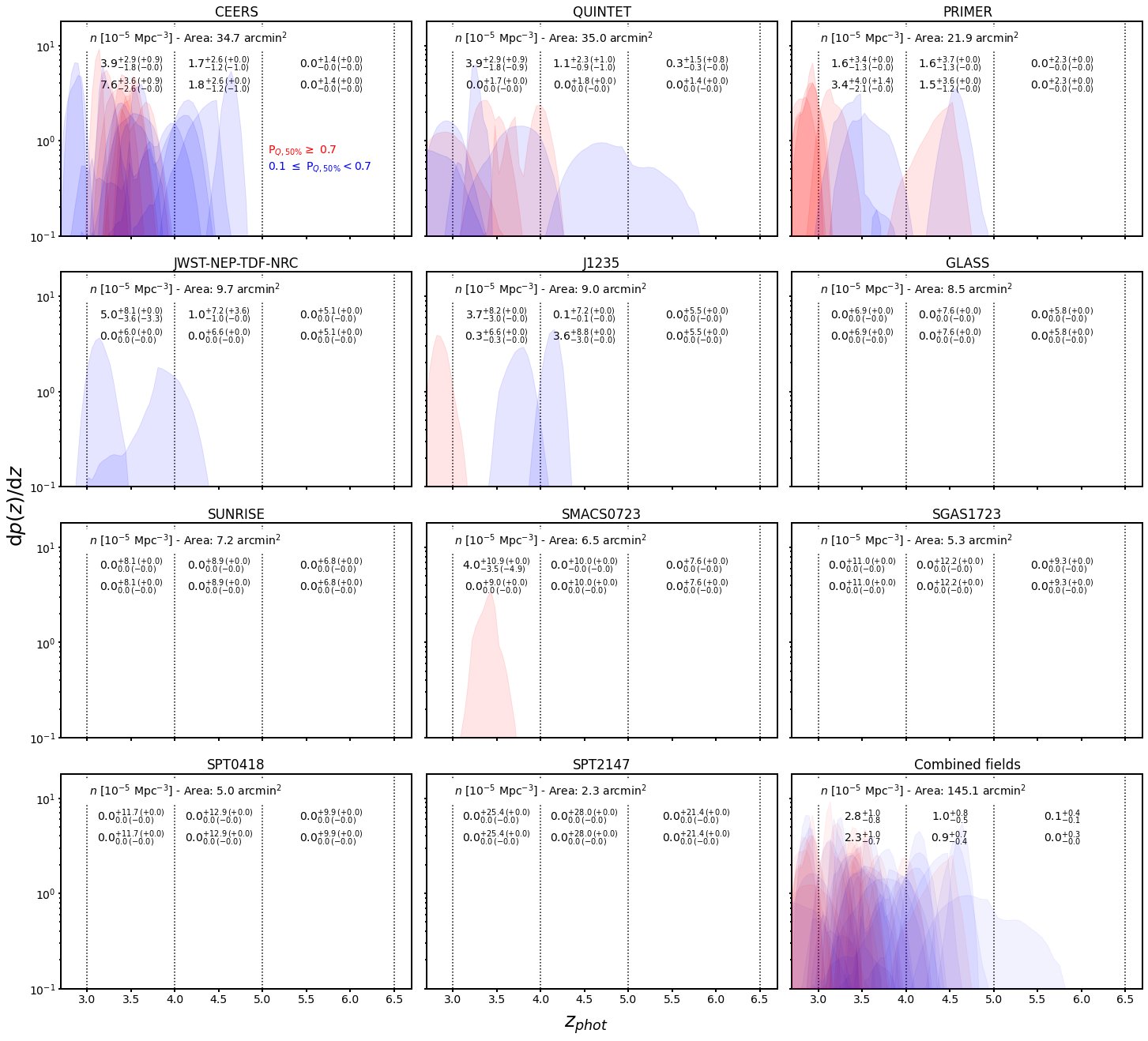}
    \caption{Number densities of $NUV-U$, $U-V$, $V-J$-selected galaxies. The red and blue areas mark the $p(z)$ of individual quiescent candidates at $M_\star \geq 10^{9.5}\,M_\odot$ with $P_{\rm Q, 50\%}\geq0.7$ and $0.1 \leq P_{\rm Q, 50\%}<0.7$, respectively. The $p(z)$ are normalized by their area ($\int_{z} p(z)\,\mathrm{d}z=1$). The sky coverage of each field and the comoving number densities per redshift bin in units of $10^{-5}$ Mpc$^{-3}$ are reported as in Figure \ref{fig:uvj_number_densities}. The first and second rows indicate $n$ in the $10^{9.5}\leq M_\star < 10^{10.6}\,M_\odot$ and $M_\star \geq 10^{10.6}\,M_\odot$ bins, respectively.}
    \label{fig:nuvuvj_number_densities}
\end{figure*}

In Figures \ref{fig:uvj_number_densities} and \ref{fig:nuvuvj_number_densities}, we show the $p(z)$ and the corresponding comoving number densities for the $UVJ$ and $NUVUVJ$-selected quiescent galaxies in each of the $11$ fields that we consider. A combined estimate based on the aggregate area of $145.1$ arcmin$^2$ is also presented. In each panel, we report the number densities in stellar mass bins of $10^{9.5}\leq M_\star < 10^{10.6}\,M_\odot$ and $M_\star \geq 10^{10.6}\,M_\odot$. The high mass threshold is chosen to directly compare these results with those in the literature (Table \ref{tab:literature}). Such a threshold also allows us to safely compare different fields. This is clear from Figure \ref{fig:mass_completeness} in Appendix \ref{appendix:mass_completeness}, showing the stellar mass limit in each field for the overall sample of galaxies and QGs. The number density estimates that we derive for the combined field are reported in Table \ref{tab:numberdensities}.\\ 

\begin{deluxetable}{lccccc}
    \tablecaption{Comoving number densities of quiescent galaxies in this work.\label{tab:numberdensities}}
    \tablehead{
    \colhead{Redshift}&
    \colhead{$\mathrm{log}(M_\star)$}&
    \colhead{$UVJ$}&
    \colhead{$UVJ$}&
    \colhead{$NUVUVJ$}&
    \colhead{$\sigma_{\rm CV}$}\\ 
    \colhead{}&
    \colhead{\small [$M_\odot$]}&
    \colhead{\small Strict}&
    \colhead{\small Padded}&
    \colhead{\small $P_{\rm Q,50\%}$}&
    \colhead{\small [\%]}
    }
    \startdata
    \multirow{2}{*}{$3 < z < 4$} & $[9.5,10.6)$& $3.9^{+1.2}_{-0.9}$& $4.1^{+1.2}_{-0.9}$& $2.8^{+1.0}_{-0.8}$& 0.10 \\
    & $>10.6$&$2.4^{+1.0}_{-0.7}$& $2.7^{+1.0}_{-0.8}$& $2.3^{+1.0}_{-0.7}$& 0.18 \\
    \rule{0pt}{4ex}\multirow{2}{*}{$4 < z < 5$} & $[9.5,10.6)$& $0.6^{+0.7}_{-0.3}$& $1.0^{+0.8}_{-0.5}$& $1.0^{+0.8}_{-0.5}$& 0.16  \\
    & $>10.6$&$0.7^{+0.7}_{-0.4}$& $0.9^{+0.8}_{-0.5}$& $0.9^{+0.7}_{-0.4}$& 0.30\\
    \rule{0pt}{4ex}\multirow{2}{*}{$5 < z < 6.5$} & $[9.5,10.6)$& $0.0^{+0.3}_{-0.0}$& $0.2^{+0.4}_{-0.2}$& $0.1^{+0.4}_{-0.1}$& 0.22 \\
    & $>10.6$&$0.0^{+0.3}_{-0.0}$& $0.0^{+0.3}_{-0.0}$& $0.0^{+0.3}_{-0.0}$& 0.41 \\
    \enddata
    \tablecomments{The comoving number densities are expressed in units of $10^{-5}$ Mpc$^{-3}$ and computed over an area of $145.1$ arcmin$^2$. The uncertainties reflect the Poissonian $1\sigma$ confidence interval. Upper limits are at $1\sigma$ using the same approach \citep{gehrels_1986}. Statistical uncertainties are accounted by integrating the $p(z)$ within the redshift intervals. The uncertainties due to cosmic variance are expressed as fractional $\sigma_{\rm CV}$ deviations (Section \ref{subsec:cv}). The selections are described in Section \ref{sec:sampleselection}. The adopted threshold for the $NUVUVJ$ selection is $P_{\rm Q,50\%}\geq 0.1$.}
\end{deluxetable}

For the $UVJ$-selected galaxies, we show the results for the sources ``strictly'' obeying the classical selection, while accounting for the color uncertainties. Adopting the ``padded'' sample returns consistent results, while estimates based on the whole ``robust'' pool of galaxies would be intended as an upper limit. An even stricter criterion accounting only for galaxies with nominal color within the standard $UVJ$ color box returns $2\times$ and $1.2\times$ lower, but fully consistent number densities in the lower and higher mass bins, respectively. In principle, an Eddington-like bias could be introduced by our ``strict'' selection coupled with the asymmetric distribution of galaxies in the color and mass space (more blue star-forming and lower mass systems can scatter into the selection box than red massive quiescent candidates that move out). However, this effect seems to be of the same order of magnitude of the statistical uncertainties. We also note that we conform to a pure color selection and we do not apply any formal correction for contamination of dusty interlopers ($\sim20$\% for standard $UVJ$ selection, \citealt{schreiber_2018c}, thus likely higher for the padded and the robust samples). The $NUVUVJ$-selected sample ($P_{\rm Q,50\%}\geq 0.1$) is as numerous as the $UVJ$ one, despite the partial overlap between the two criteria (Section \ref{subsec:overlap}), similar to what is found in \cite{gould_2022}. 

\subsection{Cosmic variance}
\label{subsec:cv}
To compute the cosmic variance, we use the prescription of \cite{steinhardt_2021}, which is based on the cookbook by \cite{moster_2011}. These authors assume a single bias parameter that links stellar to halo masses in  $\mathrm{log}(M_\star/M_\odot)$ bins of $0.5$~dex. In simulations, this assumption has been found to be valid only to a $0.2$~dex level even for massive galaxies \citep{jespersen_2022, chuang_2022}. However, this is $<1/5$ of the bin widths used in this paper and, thus, the approximation should be appropriate.
The cosmic variance is computed for each individual field taking into account the survey geometry. In order to get the cosmic variance for the bin at $\mathrm{log}(M_\star/M_\odot)=9.5-10.6$, we weight the contribution of each 0.5 dex bin to the total cosmic variance by the relative number densities in \cite{weaver_2022_smf}. The total cosmic variance is then computed as: 
\begin{equation}
    \sigma_{\rm CV,total} = \sqrt{\frac{1}{\sum_{\rm fields} \sigma_{\rm CV,field}^{-2}}}
\end{equation}

which assumes that all fields are independent.\footnote{Note that if all fields were the same shape and area, this formula reduces to the well-known $\sigma_{\rm CV,total} = \sigma_{\rm CV,field}/\sqrt{N}$}. The relative uncertainties due to cosmic variance in the combined field are reported in Table \ref{tab:numberdensities}. 
\begin{figure}
    \centering     
    \includegraphics[width=\columnwidth]{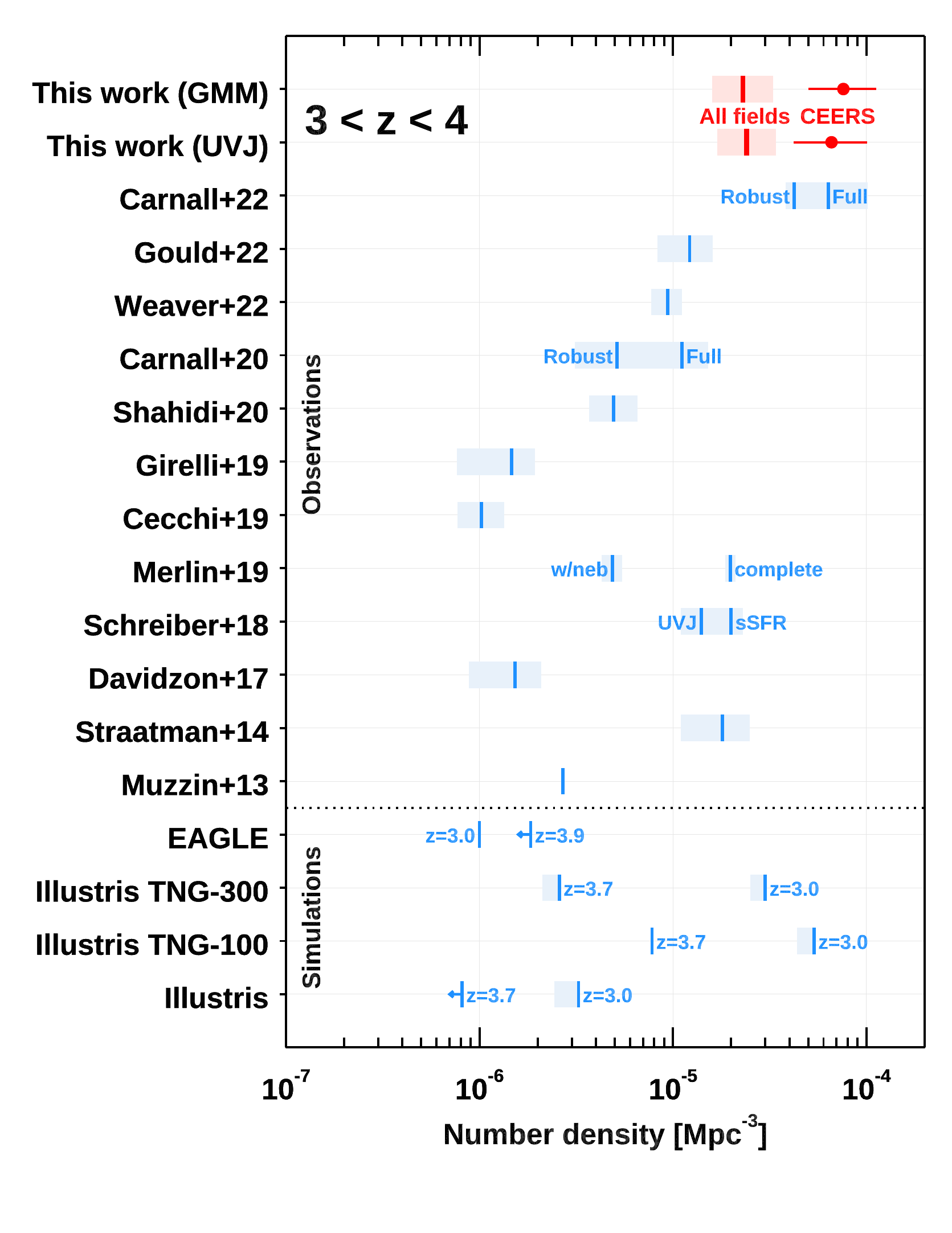}
    \caption{Comoving number densities of massive quiescent galaxies in the literature. The values have been homogenized in terms of redshift interval ($3\lesssim z\lesssim 4$) and lower mass cut ($\mathrm{log}(M_\star/M_\odot)\gtrsim10.6$, similar IMF) to the largest possible extent. The uncertainties do not include the contribution of cosmic variance. The estimates are reported in Table \ref{tab:literature} in Appendix \ref{app:literature}, along with  complementary information.}
    \label{fig:literature}
\end{figure}

\subsection{A compendium of number densities of massive quiescent galaxies at $3<z<4$}
Excellent depictions of how many QGs each individual survey or simulation find as a function of redshift are available in the literature \citep[e.g.,][]{straatman_2014, merlin_2019, girelli_2019, cecchi_2019, shahidi_2020, weaver_2022_smf, gould_2022, casey_2022, long_2022}. However, drawing direct comparisons among different works and evaluating the impact of various selections is complicated by the introduction of systematic assumptions. In Figure \ref{fig:literature}, we attempt to partially remedy this situation by reporting number densities at least adopting a consistent redshift interval ($3<z<4$) and lower mass limit for the integration ($10^{10.6}\,M_\odot$) for similar IMFs \citep{chabrier_2003, kroupa_2001}. We also add number density estimates from EAGLE \citep{schaye_2015, crain_2015}, \textit{Illustris} \citep{vogelsberger_2014}, \textit{Illustris}-TNG 100, and 300 simulations \citep{nelson_2019}. We count simulated galaxies with $\mathrm{sSFR}\leq10^{-10}$ yr$^{-1}$ within $4\times$ and $2\times$ the half-mass radius for EAGLE and \textit{Illustris}(-TNG), respectively (see \citealt{donnari_2019} for a discussion on different selection criteria of QGs, average timescales to estimate SFRs, and physical apertures in simulations). We consider snapshots at $z=3.0$ and $z=3.7-3.9$ \citep{valentino_2020a}.\\

Our number density estimates from the combined fields are of the order of $\sim2.5\times10^{-5}$ Mpc$^{-3}$, consistent with some of the determinations with similar color or sSFR cuts \citep{schreiber_2018c, merlin_2019}. Our estimates are $\sim2\times$ larger than the most recent measurements in the largest contiguous survey among those considered, COSMOS \citep{weaver_2022_smf}, also when adopting very consistent color selections \citep{gould_2022}. Interestingly, earlier determinations in the same field retrieved significantly lower estimates \citep{muzzin_2013s, davidzon_2017, girelli_2019, cecchi_2019}. This is due to a combination of deeper and homogeneous measurements in the optical and near-IR over a twice larger effective area, now available in COSMOS2020 \citep{weaver_2022}, more conservative and pure samples of QGs, the specific templates used in each work, and the integration of best-fit Schechter function underestimating the observed values at the high end of the stellar mass function.\\

The new detection and extraction based on \jwst\ LW observations allows for the selection of redder sources and improved deblending that was previously based on \hst\ bands or ground-based observations. This allows for a better identification of higher-redshift, less massive quiescent galaxies, and more robust \mstar\ by finding breaks at longer wavelengths, pinpointing objects with lower mass-to-light ratios, and removing blended objects \citep[see also the discussion in][]{carnall_2022qg}. Nevertheless, in our most massive bin, the variations among different works are still dominated by systematics in the selection, modeling, and cosmic variance.

\subsection{Field variations and groups of quiescent galaxies}
\label{subsec:field_variations}
We notice a substantial field-to-field variation especially when focusing on the most massive galaxies. Compared with the average number density on the full combined field of $\sim145$ arcmin$^2$, we find per field value oscillations of a factor of $2-3\times$ (Table \ref{tab:numberdensities_individual_field}). We ascribe these differences to cosmic variance and to the fact that massive quiescent systems might already be signpost of distant overdensities and protoclusters $-$ massive halos able to fast-track galaxy evolution. In Table \ref{tab:numberdensities_individual_field}, we report the fractional $1\sigma$ uncertainties due to cosmic variance in each field. Taken individually, an uncertainty of $\sim30-50$\% affects the number densities in the two mass bins for the largest contiguous areas that we considered. The Stephan's Quintet and CEERS fields are emblematic in this sense, appearing under- and over-dense despite a similar sky coverage. CEERS displays the largest number densities of QG $>10^{10.6}\,M_\odot$ in our compilation, consistent with the estimates in \cite{carnall_2022qg}. There we find a remarkable pair of candidate quiescent systems with consistent $z_{\rm phot}=3.54-3.38$ (\#9622-9621, also ``robust'' and not ``robust'' candidates in \citealt{carnall_2022qg}, respectively). The pair, possibly interacting, is surrounded by two more red sources with similar $z_{\rm phot}\sim3.18-3.54$ that fall in the visually vetted $UVJ$ sample (\#9490, 9329). This is reminiscent of the massive galaxies populating the ``red sequence'' in clusters, also used to find evolved protostructures at high redshift \citep{strazzullo_2015, ito_2023}. Similar QGs have already been found in overdensities at $z \gtrsim 3$ \citep{mcconachie_2022, kalita_2021, kubo_2021} or in close pairs with other massive objects \citep[``Jekyll \& Hyde'' at $z=3.717$,][of which a pair of quiescent objects would be a natural descendant]{glazebrook_2017, schreiber_2018b}. Two of these pairs or small collections of red galaxies at similar redshifts and close in projection are in our list -- not surprisingly, especially in the loosest $UVJ$-selected sample. 

\begin{figure}
    \includegraphics[width=\columnwidth]{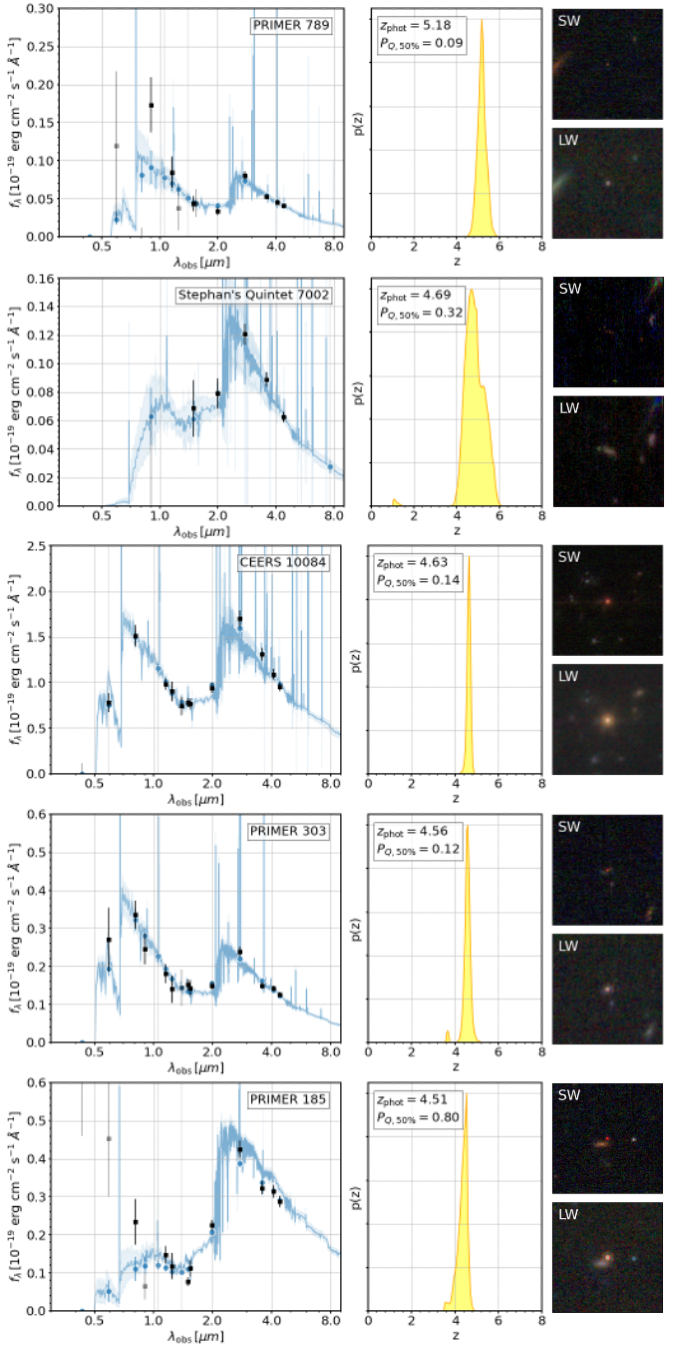}
    \caption{Robust $z>4.5$ quiescent candidates. \textit{Left column:} spectral energy distributions. Black squares and blue filled circles indicate the observed and best-fit photometry of each source. Lighter gray squares mark observed flux densities with SNR $<3$. Blue solid lines and shaded areas show the best-fit \textsc{eazy-py} models and their uncertainties. \textit{Central column:} probability distribution functions of photometric redshifts $z_{\rm phot}$ with \textsc{eazy-py}. The value of $P_{\rm Q,\,50\%}$ is reported. \textit{Right column:} SW and LW three color images of the candidates. The cutouts have sizes of $5\arcsec\times5\arcsec$.}
    \label{fig:highz_candidates}
\end{figure}

\subsection{A look at lower masses}
Figures \ref{fig:uvj_number_densities} and \ref{fig:nuvuvj_number_densities} show the number densities at lower stellar masses ($10^{9.5}\leq M_\star < 10^{10.6}$ \msun), now safely accessible with \jwst\ also at such high redshifts. In fact, the low-mass end of the \mstar\ distributions starts declining at thresholds as low as $\sim 10^{8}\,M_\odot$ at $3 < z < 6.5$ (excluding SPT2147, Figure \ref{fig:mass_completeness} in Appendix \ref{appendix:mass_completeness}). The lower limit of $M_\star=10^{9.5}\,M_\odot$ chosen for the calculation is similar to that fixed in some of the works listed in Table \ref{tab:literature}. Thus, it allows us to derive a relatively straightforward comparison, discounting some of the systematics mentioned above. At $3<z<4$, we estimate modestly higher ($\sim 1.2-1.6\times$) number densities than in the most massive bin, but in agreement within the uncertainties. The difference with previous works is similar in each bin, when available. This is consistent with the expected shape of the stellar mass function of red quiescent galaxies, roughly peaking and flattening or turning over at $\sim 10^{10.6}\,M_\odot$ at these redshifts \citep{weaver_2022_smf} and revealing a steady build up of lower-mass QGs \citep{santini_2022}. Promising low-mass QGs have been already confirmed with \jwst/NIRISS at $z\sim2.5$ in the GLASS field \citep{marchesini_2023}. We defer to future work a comprehensive analysis of the stellar mass functions of quiescent populations at these redshifts.

\subsection{High-redshift candidates}
We can now look at galaxies at $4<z<5$. As in the $3<z<4$ interval, when considering the combined fields, we estimate number densities that are $\sim2.5-4.5\times$ smaller in above and below $M_\star=10^{10.6}\,M_\odot$ than those in CEERS by \cite{carnall_2022qg}, who also perform a selection starting from \jwst\ images. The difference shrinks to a factor of $\sim2.5-1\times$ when we consider only the same field and the ``robust'' sample in their work. This seems to suggest that cosmic variance and early overdense environments are effective in producing substantial field-to-field variations also at $z>4$ (Section \ref{subsec:field_variations}). For reference, our number density estimates in the same massive bin ($M_\star\geq10^{10.6}\,M_\odot$) are consistent with those in the COSMOS field by \cite{weaver_2022_smf}, but $1.8\times$ larger than what retrieved in the same field but using a color selection similar to ours (\citealt{gould_2022}; see the discussion therein on the agreement with the latest COSMOS2020 number densities). When integrating down to lower mass limits ($10^{9.5}\,M_\odot$, but not homogenized among different works at this stage, given the impact of different depths at $z>4$), we retrieve similar $n$ as in COSMOS \citep[$(1.0\pm0.3)\times10^{-5}$ Mpc$^{-3}$ for $M_\star>10^{9.9}\,M_\odot$,][]{weaver_2022_smf} and  $1.6-2\times$ larger than in large-field \hst\ surveys such as CANDELS \citep[$\sim 7.9 \times10^{-6}$ Mpc$^{-3}$ for the ``complete'' sample at $M_\star>5\times 10^{9}\,M_\odot$,][]{merlin_2019}. Finally, the upper limits on number densities for the highest redshift bin at $5<z<6.5$ should be taken with caution, given the area covered in our analysis (Table \ref{tab:numberdensities}).\\ 

Focusing on the highest envelope of the redshift interval spanned by our sample, we find a few promising candidates at $z>4.5$. The SEDs and three-color images of the $5$ most robust sources falling either in the ``strict'' or ``padded'' $UVJ$ selections are shown in Figure \ref{fig:highz_candidates}. We do not find reliable candidates at $z\gtrsim5.2$, which signposts the earliest epoch of appearance of quiescent objects in our current sample\footnote{Two bluer and potentially quenching objects are picked by the loosest $UVJ$ selection at $z>5$. Their SEDs and color images are part of the overall release.}. Source \#185 in PRIMER ($z=4.51^{+0.16}_{-0.26}$, $\mathrm{log}(M_\star/M_\odot) = 10.9$) is also picked as among the most reliable quiescent candidates by the $NUVUVJ$ criterion ($P_{\rm Q,50\%}=0.79$). An entry with $z\sim3.2$ at $<0\farcs3$ from this source is present in previous catalogs of this field \citep{skelton_2014, mehta_2018}, but more consistent and blended with the nearby blue object (a chance projection in our analysis, $z_{\rm phot}=2.78^{+0.10}_{-0.08}$). The remaining sources are assigned lower $P_{\rm Q,50\%}$ values compatibly with their bluer colors and more recent or possibly ongoing quenching. All these candidates appear rather compact. Sources \# 789 and 303 in PRIMER are compatible with the locus of stars in the FLUX\_RADIUS, MAGAUTO plane and should be taken with a grain of salt.
For comparison, we also checked the ``robust'' objects at $z>4.5$ in \cite{carnall_2022qg}. However, we retrieve only \# 101962 (our \#2876) above $z=4$ (Appendix \ref{subappendix:carnall}, see also \citealt{kocevski_2022}). Direct spectroscopic observations with \jwst\ are necessary to break the current ceiling at $z\sim4$ imposed by atmospheric hindering to ground-based telescopes and confirm the exact redshifts of these candidates. 

\subsection{Revisiting the comparison with simulations}
For what concerns simulations, if we limit our conclusions to the homogenized massive bin and $3<z<4$ interval, we find a broad agreement with the \textit{Illustris} TNG suite at the lower end of the redshift range ($z=3$) and a rapidly increasing tension above this threshold \citep{valentino_2020a}. EAGLE and \textit{Illustris} seem to struggle to produce massive $M\geq10^{10.6}\,M_\odot$ QGs already at $z=3$, while the situation seems partially alleviated if one includes lower mass galaxies in the calculation \citep{merlin_2019, lovell_2022}. Spectroscopically confirmed massive QG at $z>4$ would not only exacerbate the tension with these simulations, but also for the latest-generation examples, such as FLARES \citep{lovell_2021, vijayan_2021}. We do not find any $M_\star>10^{10.6}\,M_\odot$ objects with $\mathrm{sSFR}<10^{-10}\,\mathrm{yr}^{-1}$ in EAGLE or \textit{Illustris}-TNG at $z=4$ and above, while FLARES produces $2$~dex lower number densities at $z=5$ ($n=7.2\times10^{-8}\,\mathrm{Mpc}^{-3}$, \citealt{lovell_2022}).

\section{Conclusions}
\label{sec:conclusions}
We present a sample of $\sim80$ \jwst-selected candidate quiescent and quenching galaxies at $z>3$ in 11 separate fields with publicly available imaging collected during the first 3 months of telescope operations. We homogeneously reduce the \jwst\ data and combine them with available \hst\ optical observations. We both perform a classical $UVJ$ selection and apply a novel technique based on Gaussian modeling of multiple colors -- including an $NUV$ band sensitive to recent star formation, which is necessary to explore the quenching of galaxies in the early Universe. Here we focus on a basic test for simulations and empirical models: the estimate of comoving number densities of this population. 
\begin{itemize}
    \item We estimate $n\sim2.5\times10^{-5}$ Mpc$^{-3}$ for massive candidates ($\geq 10^{10.6}\,M_\odot$) with both selections, but substantial field-to-field variations of the order of $2-3\times$. This is likely due to cosmic variance ($\sim30-50\%$ uncertainty in the largest contiguous fields of $\sim30\,\mathrm{arcmin}^2$ such as CEERS or Stephan's Quintet) and the fact that early and evolved galaxies might well trace matter overdensities and the emerging cores of protoclusters already at $z>3$. We find promising candidate pairs or groups of quiescent or quenching galaxies with consistent redshifts in the field with the highest number density.
    \item We compile and homogenize the results of similar attempts to quantify the number densities of massive QG at $3<z<4$ in the literature. The comparison across almost $20\times$ different determinations highlights the impact of cosmic variance and systematics primarily in the selection techniques. The most recent estimates seem to converge toward a value of $n\sim1-2\times10^{-5}$ Mpc$^{-3}$ -- not exceedingly far from what established via ground-based observations.
    \item We apply our homogenization also to publicly accessible large-box cosmological simulations. As noted in the literature, a tension with observations at increasing redshifts is evident -- to the point that even a single confirmation of a massive QG at $z~>~4~-~4.5$ would challenge some of the theoretical predictions. A handful of promising candidates up to $z\sim5$ are found in our systematic search and presented here.
    \item We start exploring the realm of lower mass QG candidates, taking advantage of the depth and resolution of \jwst\ at near-IR wavelengths. We measure number densities at $10^{9.5} \leq M_\star < 10^{10.6}\,M_\odot$ similar to those at $\geq10^{10.6}\,M_\odot$, consistent with the expected flattening or turnover of the stellar mass function of quiescent objects and the onset of the low-mass quenched population.
\end{itemize}
This work is the first of a series of articles that will focus on the characterization of several aspects of the sample selected here (morphologies, SED and SFH modeling, also resolved, and environment). We remark that all of the high-level science products (notably catalogs, images, and SED best-fit parameters) are publicly available. The continuous flow of new \jwst\ imaging data and -- soon -- systematic spectroscopic coverage of large portions of the sky (e.g., Cosmos-Web, \citealt{casey_2022}; UNCOVER, \citealt{bezanson_2022}; GO 2665, PI: K. Glazebrook: \citealt{nanayakkara_2022}; 2362: PI: C. Marsan; 2285, PI: A. Carnall) will allow us to shrink the uncertainties due to cosmic variance and pursue the research avenues highlighted throughout the manuscript, starting with the necessary spectroscopic confirmation.

\acknowledgments
We acknowledge the careful reading and the constructive comments from the anonymous referee. We warmly thank Emiliano Merlin, Giacomo Girelli, Abtin Shahidi, and Micol Bolzonella for computing and sharing their number densities in the specified redshift and mass intervals. We also thank Dan Coe for sharing the magnification maps computed by the RELICS team. This work is based on observations made with the NASA/ESA/CSA \textit{James Webb Space Telescope}. The data were obtained from the Mikulski Archive for Space Telescopes at the Space Telescope Science Institute, which is operated by the Association of Universities for Research in Astronomy, Inc., under NASA contract NAS 5-03127 for JWST. The specific observations analyzed can be accessed via \dataset[10.17909/g3nt-a370]{https://doi.org/10.17909/g3nt-a370}. These observations are associated with programs ERS \#1324, 1345, and 1355; ERO \#2736; GO \#1837 and 2822; GTO \#2738; and COM \#1063. The authors acknowledge the teams and PIs for developing their observing program with a zero-exclusive-access period.
The Cosmic Dawn Center (DAWN) is funded by the Danish National Research Foundation under grant No. 140.
S.F. acknowledges the support from NASA through the NASA Hubble Fellowship grant HST-HF2-51505.001-A awarded by the Space Telescope Science Institute, which is operated by the Association of Universities for Research in Astronomy, Incorporated, under NASA contract NAS5-26555.
M.H. acknowledges funding from the Swiss National Science Foundation (SNF) via a PRIMA Grant PR00P2 193577 ``From cosmic dawn to high noon: the role of black holes for young galaxies''. 
This work was supported by JSPS KAKENHI Grant Numbers JP21K03622, 20K14530, and 21H044902. 
K.I. acknowledges support from JSPS grant 22J00495.
G.E.M. acknowledges the Villum Fonden research grants 13160 and 37440.
O.I. acknowledges the funding of the French Agence Nationale de la Recherche for the project iMAGE (grant ANR-22-CE31-0007).
\vspace{5mm}
\facilities{\textit{JWST, HST}}

\software{\textsc{grizli} \citep{brammer_2021, brammer_2022}, 
          \textsc{eazy-py} \citep{brammer_2008}, 
          \textsc{sep} \citep{barbary_2016},  
          \textsc{astropy} \citep{astropy_2022},  
          \textsc{astrodrizzle} \citep{fruchter_2002, koekemoer_2003},
          \textsc{glafic} \citep{oguri_2010, oguri_2021b}.
          }

\newpage
\appendix
\renewcommand{\thefigure}{\thesection.\arabic{figure}}
\setcounter{figure}{0}

\begin{deluxetable*}{lll}
    \tablecaption{Filter coverage in each field.\label{tab:coverage}}
    \tabletypesize{\normalsize}
    \tablehead{
    \colhead{Field}&
    \colhead{\jwst\ wavelength}&
    \colhead{\hst\ wavelength}\\
    \colhead{}&
    \colhead{[$\mu$m]}&
    \colhead{[$\mu$m]}
    }
    \startdata
         CEERS             & 1.15, 1.5, 2, 2.7, 3.5, 4.1\tablenotemark{a}, 4.4 &
                             0.6, 0.8, 0.44, 1.05, 1.25, 1.4, 1.6 \\
         Stephan's Quintet & 1.5, 2, 2.7, 3.5, 4.4 & 
                            -- \\
         PRIMER            & 0.9, 1.15, 1.5, 2, 2.7, 3.5, 4.1\tablenotemark{a}, 4.4 &
                             0.44, 0.6, 0.8, 1.05, 1.25, 1.4, 1.6 \\
         NEP               & 0.9, 1.15, 1.5, 2, 2.7, 3.5, 4.1\tablenotemark{a}, 4.4 &
                             0.44, 0.6 \\    
         J1235             & 0.7, 0.9, 1.15, 1.5, 2, 2.7, 3.0\tablenotemark{a}, 3.5, 4.4, 4.8\tablenotemark{a} &
                            --\\
         GLASS             & 0.9, 1.15, 1.5, 2, 2.7, 3.5, 4.4 &
                             0.44, 0.48, 0.6, 0.78, 0.8, 1.05, 1.25, 1.4, 1.6\\
         Sunrise           & 0.9, 1.15, 1.5, 2, 2.7, 3.5, 4.1\tablenotemark{a}, 4.4 & 
                             0.44, 0.48, 0.6, 0.8, 1.05, 1.10, 1.25, 1.4, 1.6\\
         SMACS0723         & 0.9, 1.5, 2, 2.7, 3.5, 4.4 &
                             0.44, 0.6, 0.8, 1.05, 1.25, 1.4, 1.6\\
         SGAS1723          & 1.15, 1.5, 2, 2.7, 3.5, 4.4 & 
                             0.48, 0.6, 0.78, 0.8, 1.05, 1.10, 1.4, 1.6\\
         SPT0418           & 1.15, 1.5, 2, 2.7, 3.5, 4.4 &
                             --\\
         SPT2147           & 2, 2.7, 3.5, 4.4 & 
                            1.4 \\
    \enddata
    \tablecomments{\textbf{\jwst\ NIRCam filter identifiers:} Wide (W): $0.7=\mathrm{F070W}$; $0.9=\mathrm{F090W}$; $1.15=\mathrm{F115W}$; $1.5=\mathrm{F150W}$; $2=\mathrm{F200W}$; $2.77=\mathrm{F277W}$; $3.5=\mathrm{F356W}$; $4.4=\mathrm{F444W}$; Medium (M): $3.0=\mathrm{F300M}$; $4.1=\mathrm{F410M}$; $4.8=\mathrm{F480M}$. \textbf{\hst\  filter identifiers:} $0.44=\mathrm{ACS/F435W}$; $0.48=\mathrm{ACS/F475W}$; $0.6=\mathrm{ACS/F606W}$; $0.78=\mathrm{ACS/F775W}$; $0.8=\mathrm{ACS/F814W}$; $1.05=\mathrm{WFC3/F105W}$; $1.25=\mathrm{WFC3/F125W}$; $1.4=\mathrm{WFC3/F140W}$; $1.6=\mathrm{WFC3/F160W}$}
    \tablenotetext{a}{Medium-band filter.}
\end{deluxetable*}

\section{Fields}
\label{appendix:fields}
\begin{figure*}
    \includegraphics[width=\textwidth,height=\textheight,keepaspectratio]{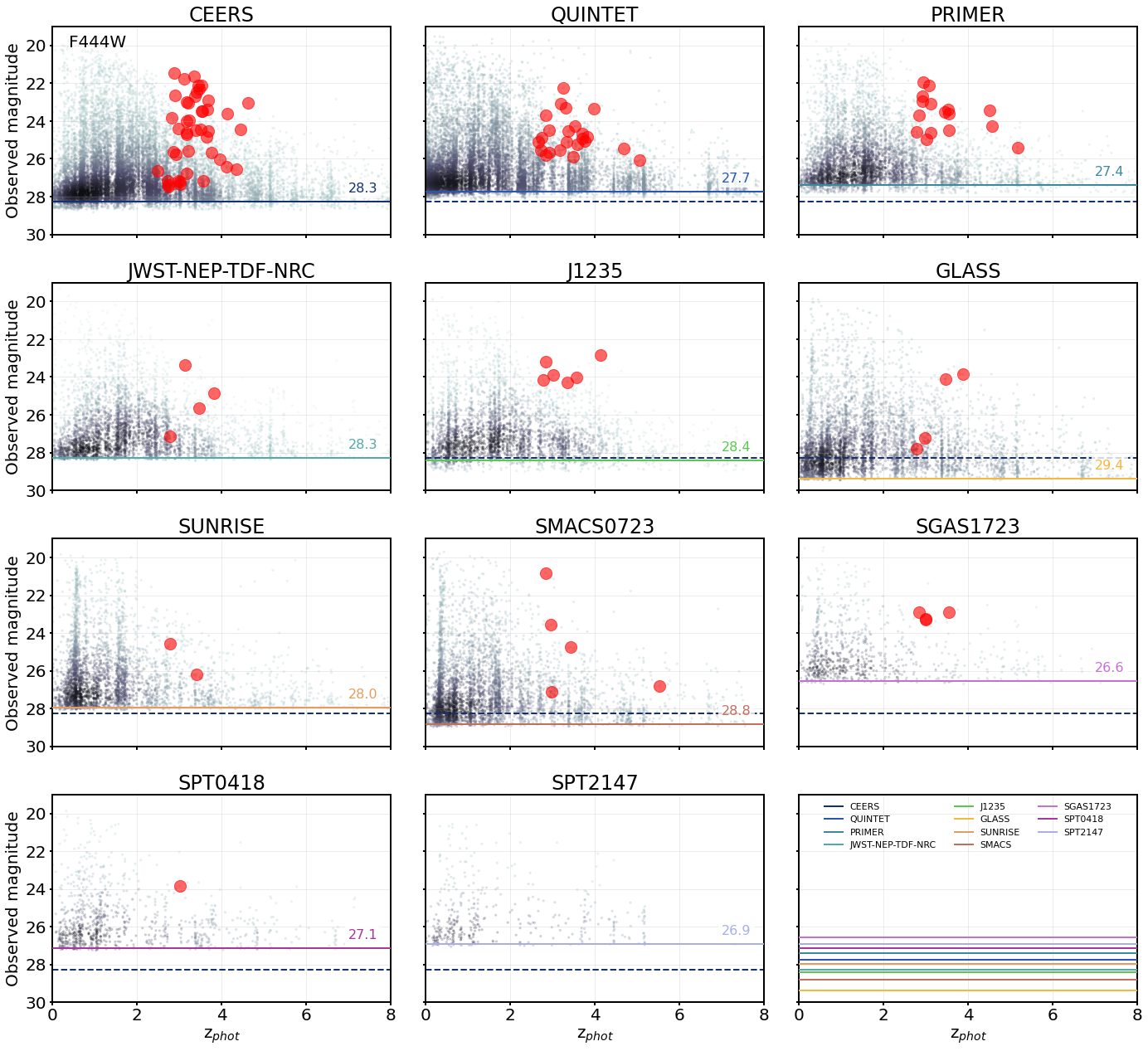}
    \caption{Observed NIRCam F444W magnitudes as a function of the photometric redshifts. Gray points indicate sources in each field as label. The color intensity scales as the density of points. Red circles show our $UVJ$-selected sample of quiescent candidates at $3<z<6.5$ after the visual inspection. The color lines mark the $5\sigma$ depths in $0\farcs5$ diameter apertures in F444W. For reference, we show the depth for the CEERS field in each panel (dashed blue line). A direct comparison of the depth is shown in the bottom right panel.}
    \label{fig:f444w_limits}
\end{figure*}

Here we provide a brief summary of the available observations for each field. The \jwst\ and \hst\ imaging availability is summarized in Table \ref{tab:coverage}. Figure \ref{fig:f444w_limits} shows the depths in F444W within the apertures used for the photometric extraction. Similar plots for F150W, F200W, F277W, and F356W are available in Figure Set A1. The depths are reported in Table \ref{tab:table_obs}. The minimum overlap of the NIRCam bands imposed for the selection and corresponding to the areas in Table \ref{tab:coverage} is shown in Figure \ref{fig:coverage}. 

\subsection{CEERS}
\label{appendix:fields:ceers}
The Cosmic Evolution Early Release Science Survey (CEERS) is among the Director Discretionary Early Release Science (DD-ERS) programs (ERS 1345, PI: S. Finkelstein). It targeted the Extended Groth Strip (EGS) \hst\ legacy field with several \jwst\ instruments for imaging and, in the future, spectroscopy (\citealt{bagley_2022} for a full description of the program and an official data release of the CEERS team). In this work, we made use of the NIRCam imaging in the ``wide'' F115W, F150W, F200W, F277W, F356W, and F444W filters, plus the ``medium" F410M band. We incorporated available \hst\ observations from the archive \citep[CHArGE,][]{kokorev_2022}. 

\subsection{Stephan's Quintet}
\label{appendix:fields:quintet}
\begin{figure*}
    \includegraphics[width=0.5\textwidth]{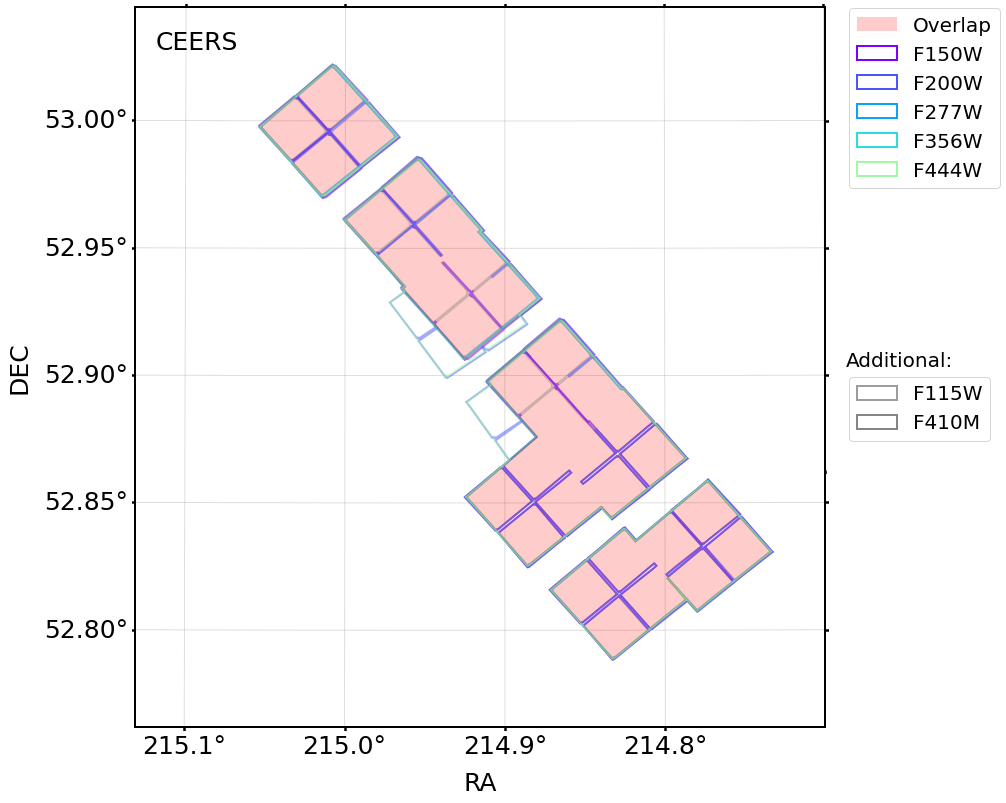}
    \includegraphics[width=0.5\textwidth]{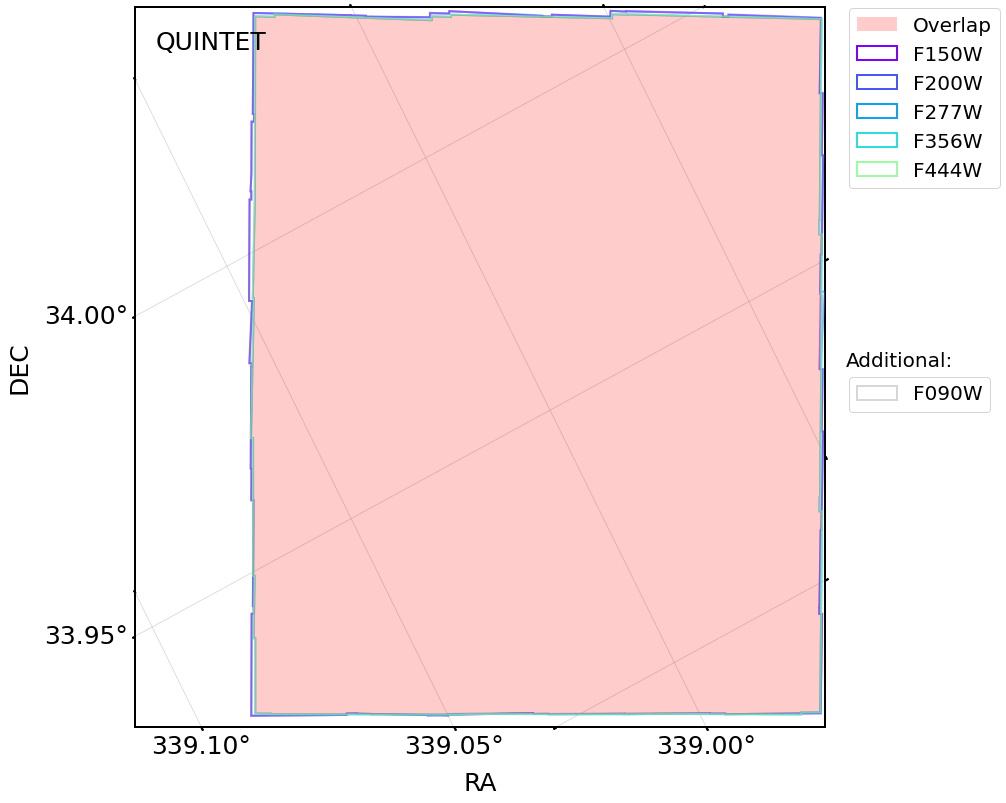}
    \includegraphics[width=0.5\textwidth]{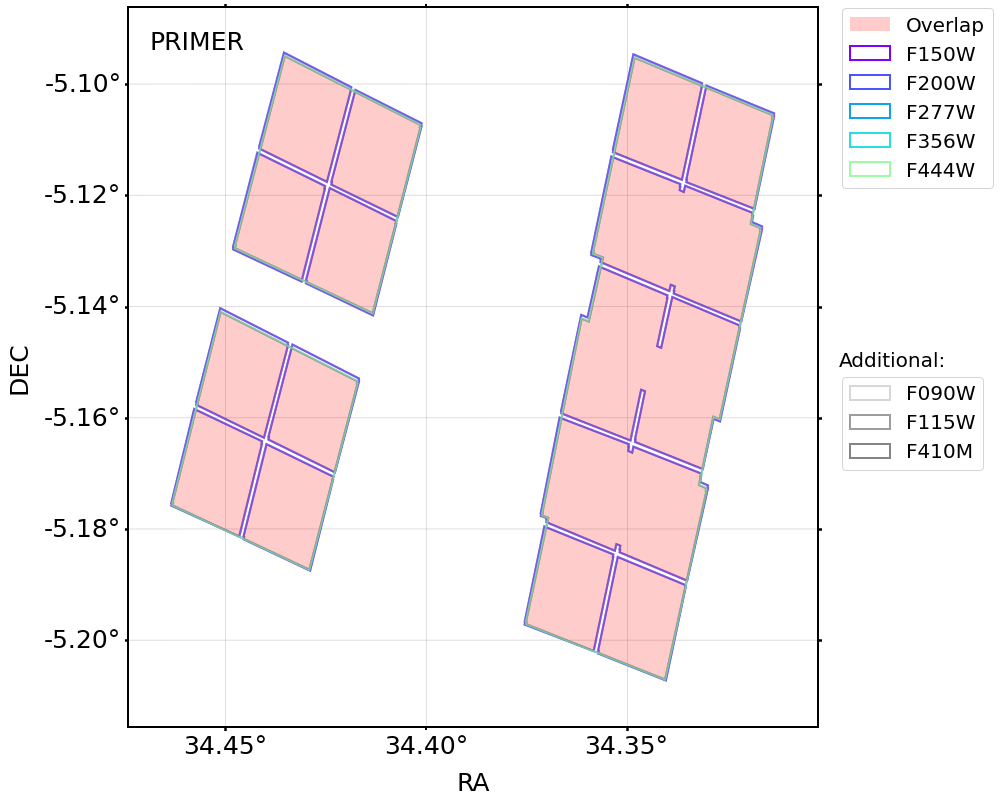}
    \includegraphics[width=0.5\textwidth]{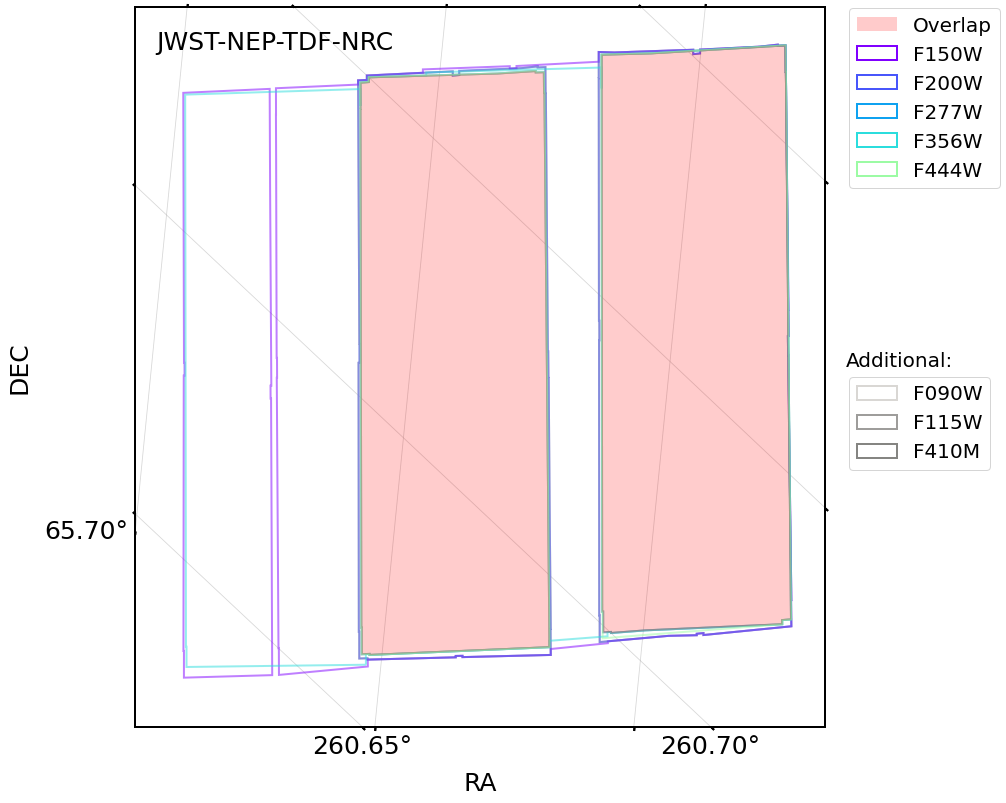}
    \includegraphics[width=0.5\textwidth]{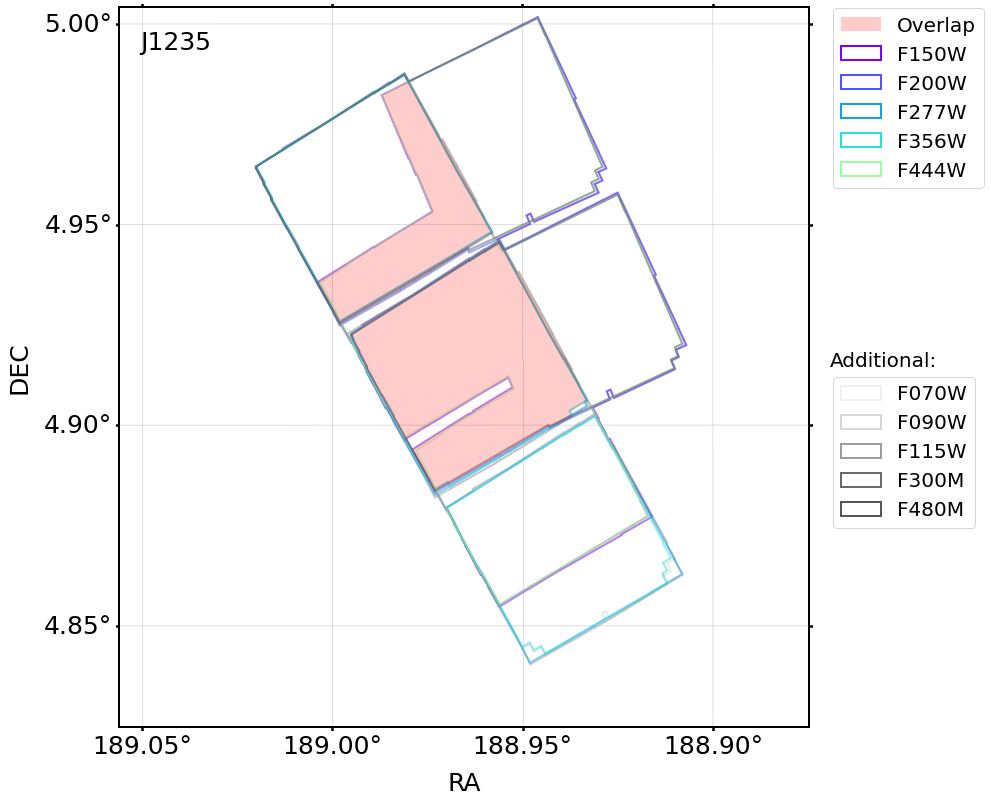}
    \includegraphics[width=0.5\textwidth]{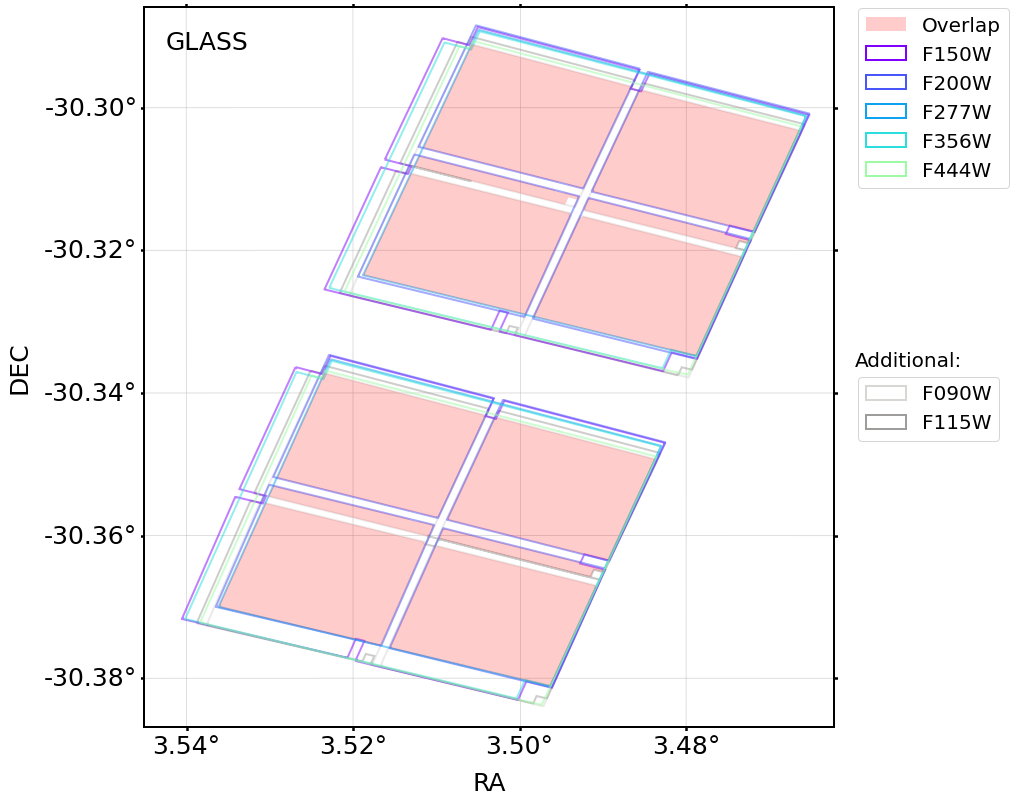}
       
    \caption{\jwst\ coverage maps. For every field, we show the footprint of each \jwst\ filter colored as labeled. The red shaded area indicates the overlap of our selection filters (F150W, F200W, F277W, F356W, and F444W).}
    \label{fig:coverage}
\end{figure*}

 Stephan's Quintet has been targeted and the images immediately released as part of the Early Release Observations \citep[ERO \# 2736,][]{pontoppidan_2022}. No \hst\ coverage is available in our archive. In Figure \ref{fig:coverage}, we show the nominal overlap of the filters that we required for the selection, but we carved a large portion of the central part of the field where contamination from the galaxies belonging to the group was too high to ensure good quality photometry. This effectively reduced the area by $\sim5$ arcmin$^2$. 

\subsection{PRIMER}
\label{appendix:fields:primer}
The Public Release IMaging for Extragalactic Research (PRIMER, GO 1837, PI: J. Dunlop) is a Cycle 1 accepted program targeting contiguous areas in the COSMOS \citep{scoville_2007} and Ultra-Deep Survey \citep[UDS,][]{lawrence_2007} fields with NIRCam and MIRI. Here we considered the area covered with NIRCam in the UDS field, critically overlapping with the \hst\ deep imaging from the Cosmic Assembly Near-infrared Deep Extragalactic Legacy Survey \citep[CANDELS,][]{grogin_2011}.

\subsection{North Ecliptic Pole}
\label{appendix:fields:nep}
The North Ecliptic Pole (NEP) Time-Domain Field (TDF) is being observed as part of the Guaranteed-Time Observations (GTO, program 2738, PI: R. Windhorst). The first spoke in the TDF was immediately released to the public. Here we considered the portion of the sky observed by NIRCam. Coverage with \hst\ ACS/F435W and F606W is available.

\subsection{J1235}
\label{appendix:fields:j1235}
J1235 is a low-ecliptic latitude field observed during commissioning with the largest compilation of wide and medium NIRCam filters in our collection in Cycle 0 (COM/NIRCam 1063, PI: B. Sunnquist). The goal was to verify to a 1\% accuracy the flat fielding after launch and to accumulate sky flats for future calibration programs. No \hst\ imaging available.

\subsection{GLASS Parallel}
\label{appendix:fields:glass}
Parallel NIRCam observations were acquired while observing Abell 2744 as part of the DD-ERS program ``GLASS-JWST'' (ERS 1324, PI: T. Treu, \citealt{treu_2022}). The parallel fields are sufficiently far from the cluster that gravitational lensing does not appreciably affect our work. Abell 2744 has been targeted by several \hst\ programs, including the Grism Lens-Amplified Survey from Space (GLASS) itself and a project tailored to maximally exploit the scientific return of the parallel fields (GO/DD 17231, PI: T. Treu), which we also included in our data.

\subsection{Sunrise}
\label{appendix:fields:sunrise}
We dubbed the cluster-lensed field WHL0137-08 from the Reionization Lensing Cluster Survey \citep[RELICS,][GO \#2822]{coe_2019} after the ``Sunrise arc" that was discovered in it \citep{salmon_2020, vanzella_2022} -- even hosting a highly magnified star \citep{welch_2022}. Being part of RELICS, ample \hst\ ancillary data is available. The area reported in Table \ref{tab:table_obs} accounts for the lensing effect at $z=3-5$. \jwst\ data were included in updated magnification maps generated with \textsc{glafic} \citep{oguri_2010, oguri_2021b}.

\subsection{SMACS0723}
\label{appendix:fields:smacs0723}
SMACS0723 is also part of the RELICS survey, one of the spectacular, and immediately released ERO objects \citep{pontoppidan_2022}. Here we made use of NIRCam imaging from detectors targeting the cluster and a position offset from it. MIRI, when available, was included. Also in this case we accounted for the effect of lensing in the area centered on the cluster using an updated version of previous magnification maps with \textsc{glafic}, now including \jwst\ data. \hst\ coverage from RELICS is available. Long-wavelength observations from the ALMA Lensing Cluster Survey (PI: K. Kohno) were used to look for possible dusty contaminants, when available.

\subsection{SGAS1723, SPT0418, and SPT2147}
\label{appendix:fields:templates}
These are fields from the Targeting Extremely Magnified Panchromatic Lensed Arcs and Their Extended Star formation (TEMPLATES) ERS program (ERS 1355, PI: J. Rigby). The primary targets are 4 strongly galaxy-lensed systems with ample ancillary data across the electromagnetic spectrum. In this work, we relied on the imaging portion of the ERS program. Single galaxy lensing does not affect the field on large scales. SPT2147 was not imaged with the F115W and F150W filters.

\setcounter{figure}{1}
\begin{figure*}
    \includegraphics[width=0.5\textwidth]{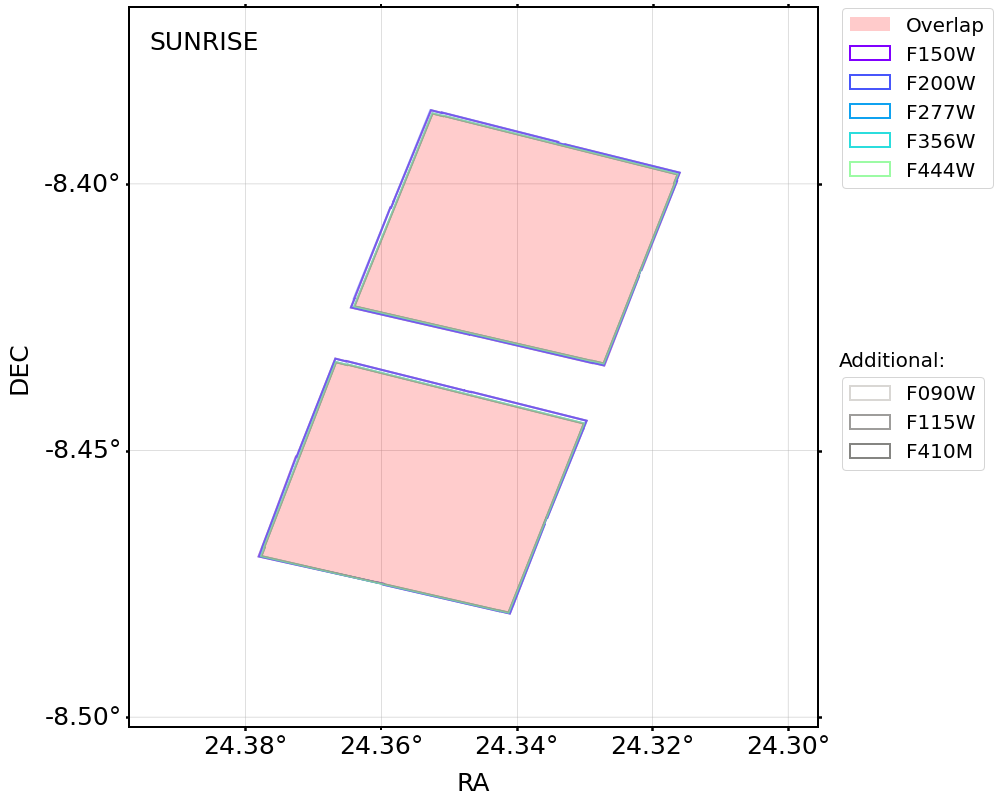}
    \includegraphics[width=0.5\textwidth]{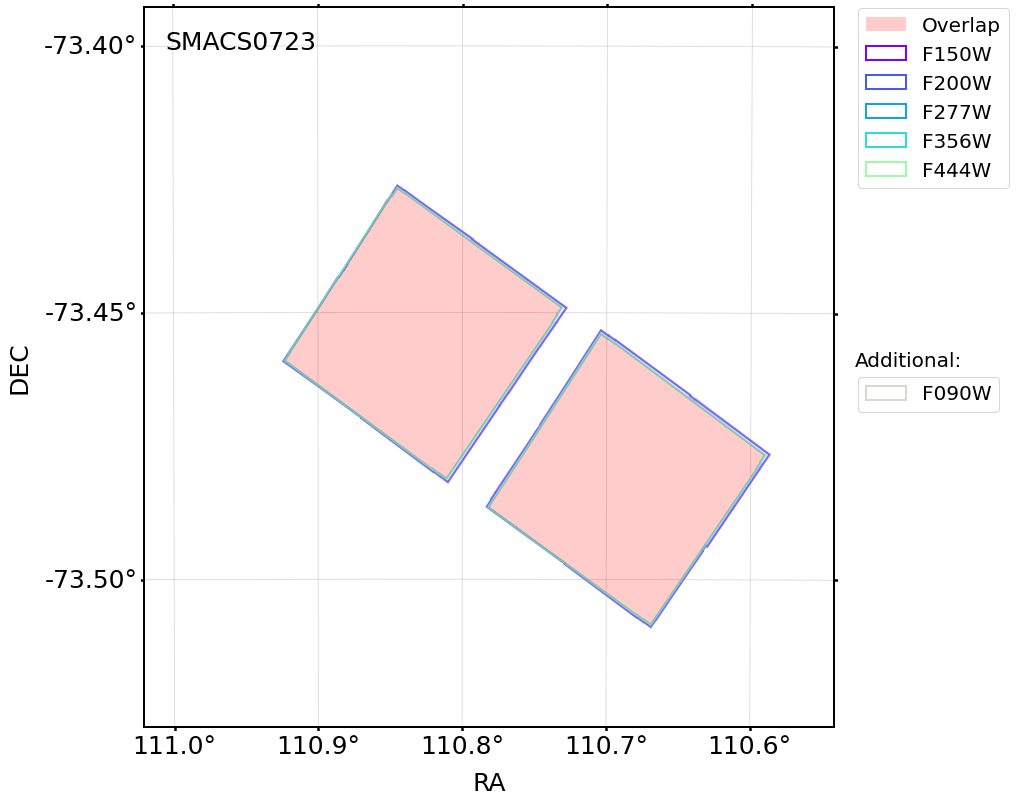}
    \includegraphics[width=0.5\textwidth]{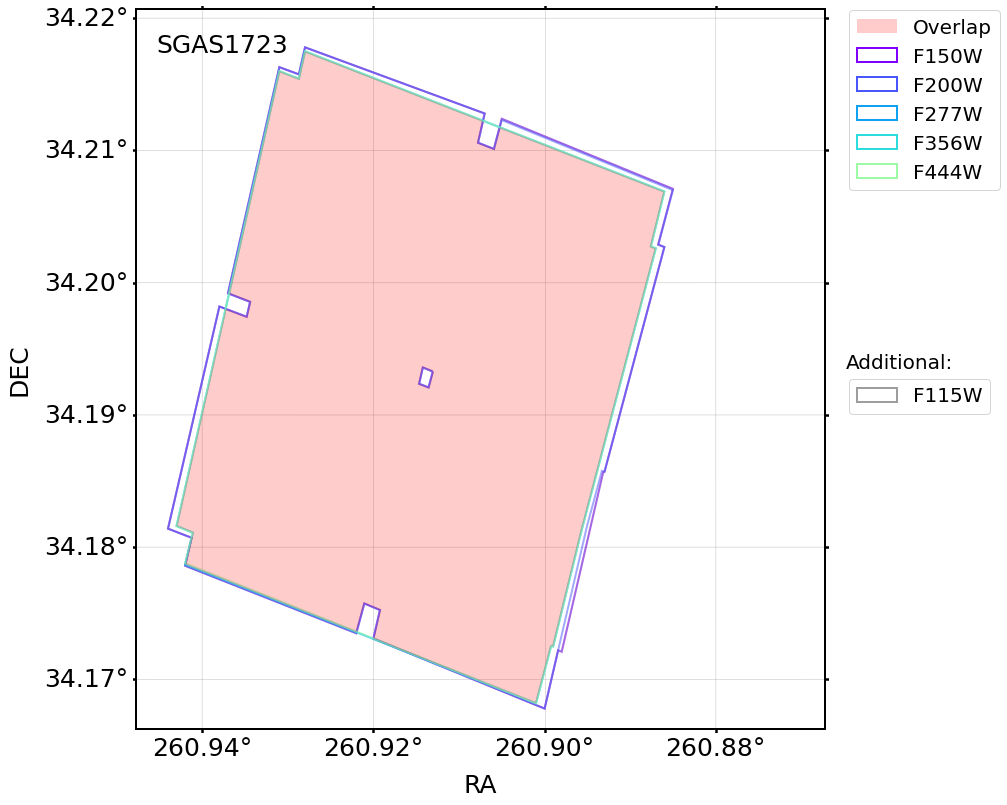}
    \includegraphics[width=0.5\textwidth]{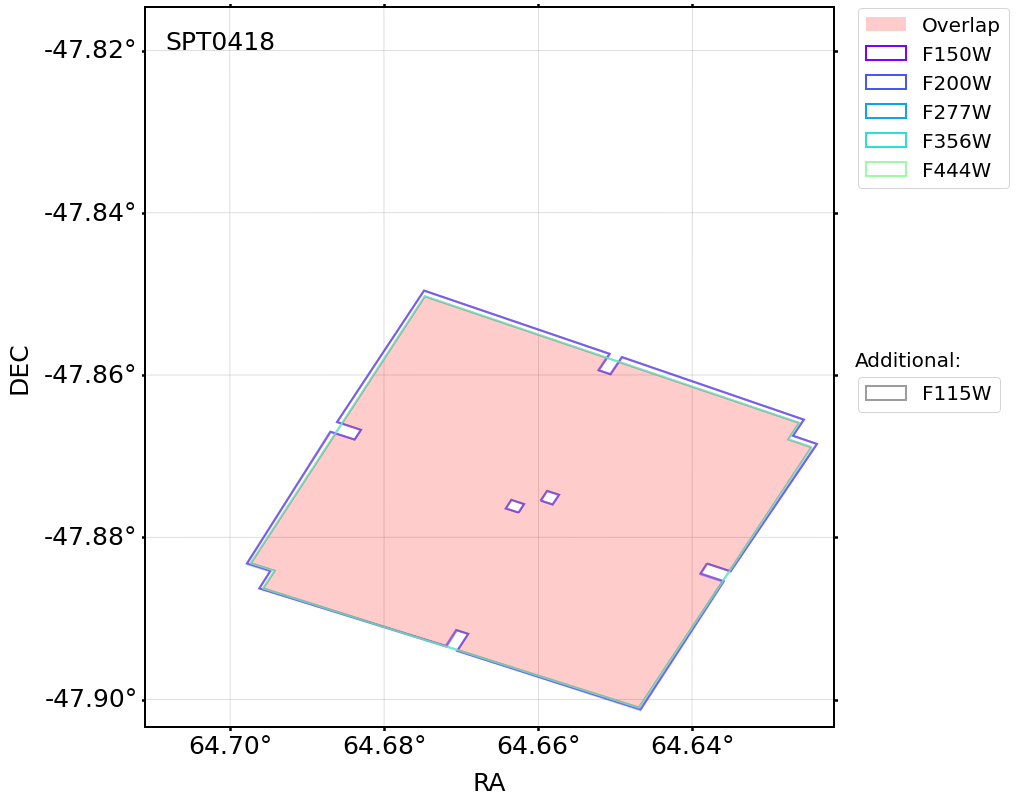}
    \includegraphics[width=0.5\textwidth]{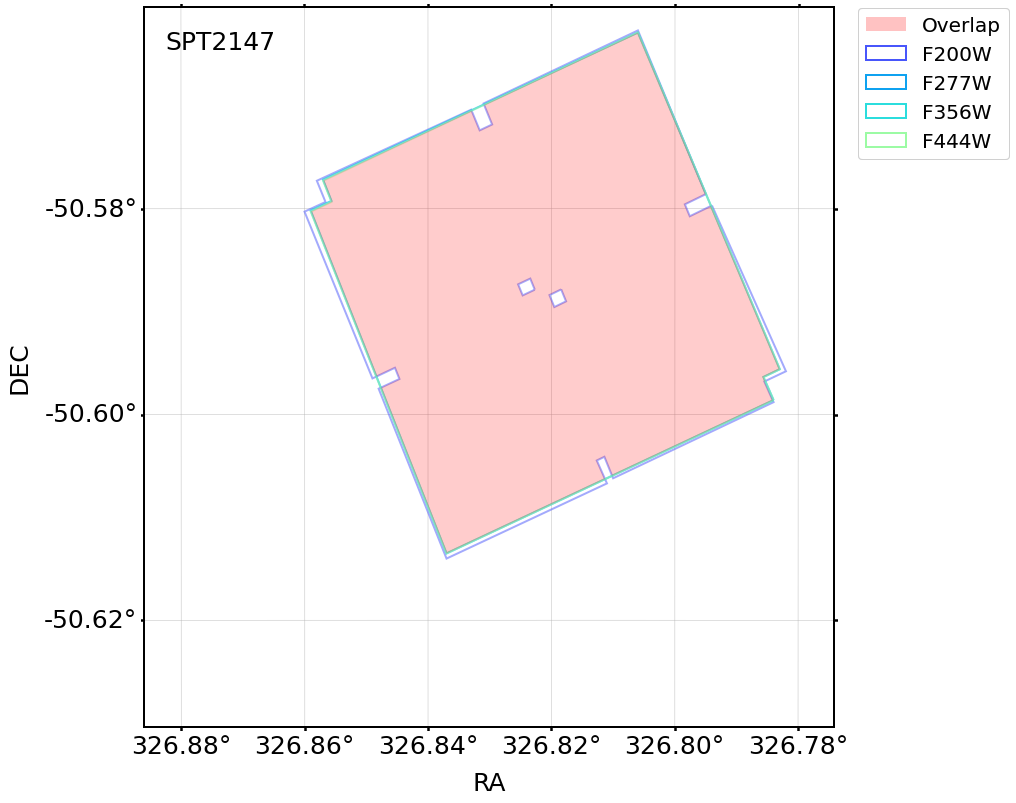}
    \caption{(Continue.)}
\end{figure*}

\section{Sanity checks on the sample}
\label{appendix:checks}
Here we describe in more detail the tests on the robustness of our sample selection to which we briefly referred in Section \ref{subsec:sanity}.

\subsection{Photometry and source extraction}
\label{subappendix:3dhst}
\begin{figure*}
    \centering
    \includegraphics[width=\textwidth]{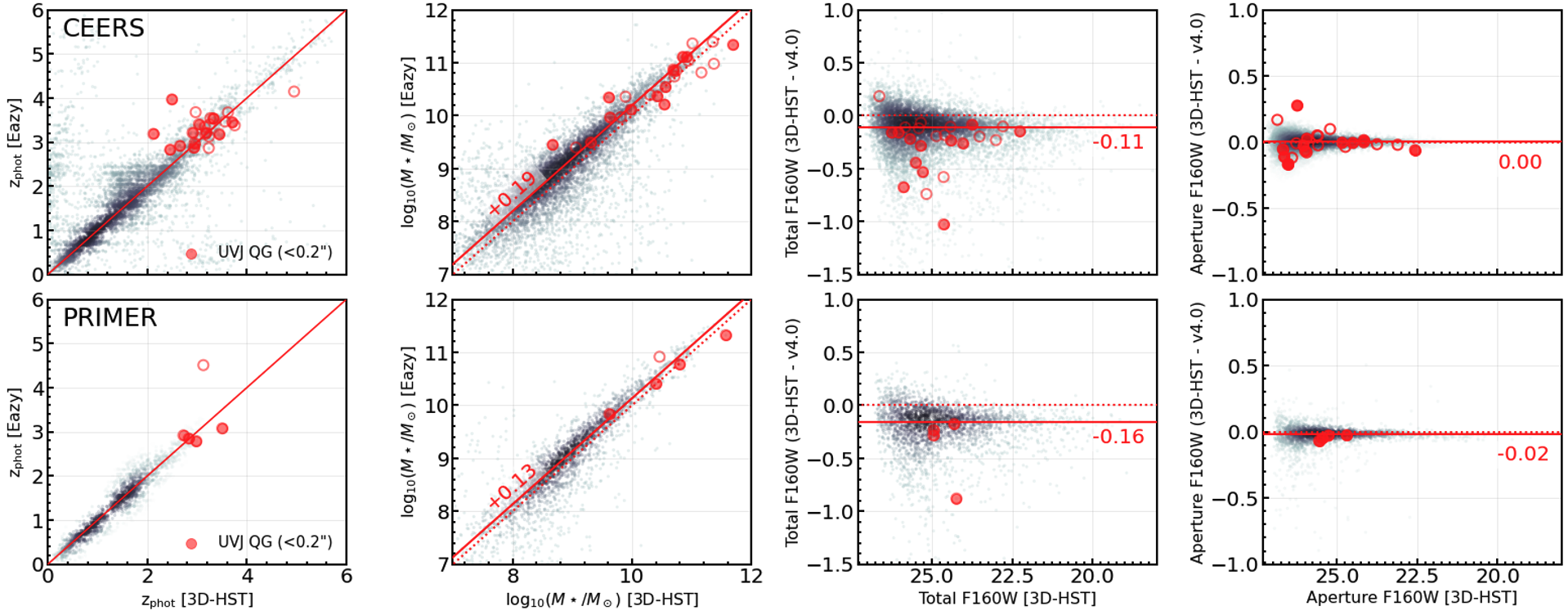}
    \caption{Comparison with 3D-\hst. Gray points indicate sources in common (maximal separation $<0\farcs5$) between our catalogs in CEERS (top row) and PRIMER (bottom row) and those in 3D-\hst\ from \cite{skelton_2014}. The color intensity scales as the density of points. Red filled and empty circles mark $UVJ$-selected QGs from our sample with a counterpart at $<0\farcs2$ and $<0\farcs5$ in 3D-\hst, respectively. From the left to right: photometric redshifts; stellar masses; total photometry in \hst/F160W (in our analysis derived from the reference $0\farcs5$ aperture); photometry in the same band in a common $0\farcs7$ diameter aperture. The median offsets from the one-to-one relation (dotted lines) are shown, when applicable.}
    \label{fig:3dhst}
\end{figure*}

We compared our photometric extraction and SED modeling (Section \ref{sec:data}) with those from 3D-\hst\ \citep{skelton_2014} for sources in CEERS (EGS) and PRIMER (UDS). We matched sources allowing for a maximal $<0\farcs5$ separation. Figure \ref{fig:3dhst} shows the comparison in $z_{\rm phot}$, \mstar, total \hst/F160W photometry computed from our reference $0\farcs5$ diameter aperture, and that within a common $0\farcs7$ aperture. The agreement is overall excellent, despite different detection images and corrections applied. The aperture photometry is fully consistent and so are the $z_{\rm phot}$ estimates from the previous and current version of \textsc{eazy(-py)}. The F160W total magnitudes computed starting from the $0\farcs5$ diameter apertures considered here are fainter than those from $0\farcs7$ in 3D-\hst\ in CEERS and PRIMER: the median differences are $0.11\,(\sigma_{\rm MAD}=0.14)$ and $0.16\,(0.14)$ mag, respectively. However, at fixed aperture, the total magnitudes are fully consistent. The difference arises from the  detection bands and where the aperture correction is computed: F160W for 3D-\hst\ and the NIRCam combined LW image in our analysis. Finally, the total \mstar\ are systematically lower in 3D-\hst\ than in our \jwst-based catalogs of CEERS and PRIMER, with median differences and MAD of $0.19\,(\sigma_{\rm MAD}=0.30)$ and $0.13\,(0.24)$ dex, respectively. All things considered, the offsets are fully ascribable to the different recipes adopted to estimate these quantities and consistent with typical systematic uncertainties inevitably present when we compare different catalogs of the same sources. Our samples of $UVJ$ and $NUVUVJ$ selected QG at $z>3$ do not appreciably deviate from these trends.   

\subsection{Availability of \hst\ photometry}
\label{subappendix:filters}
\begin{figure*}
    \includegraphics[width=\textwidth,height=\textheight,keepaspectratio]{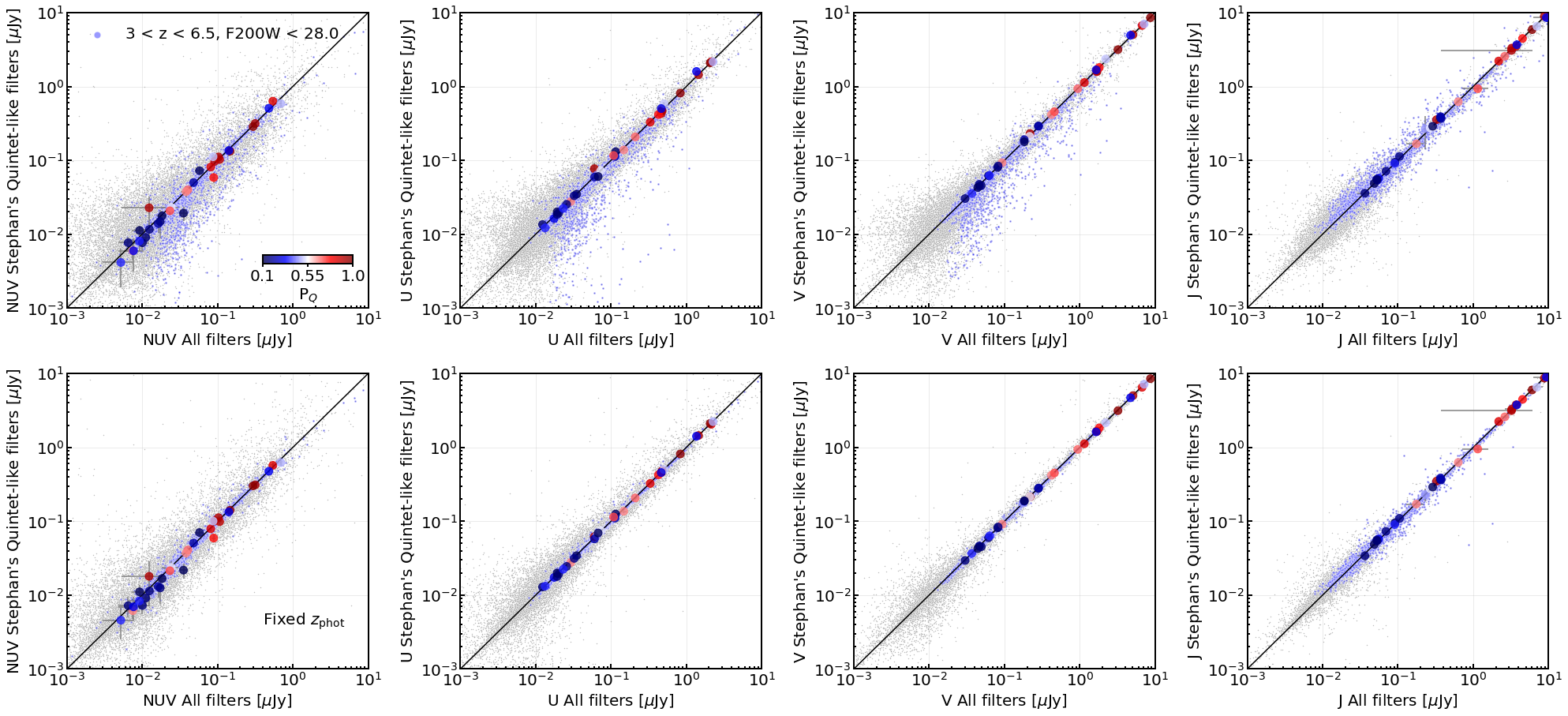}
    \caption{Effect of \hst\ filter coverage. From left to right: rest-frame $NUV$, $U$, $V$, and $J$ flux densities from SED modeling with \textsc{eazy-py} in the CEERS field. Quantities on the X-axes are computed with the full filter set, including \hst\ bands; those on the Y-axes are from the run with NIRCam wide filters only. The flux densities are computed at their respective best-fit $z_{\rm phot}$ in the top row and at fixed $z_{\rm phot}(\rm full\,filter\,set)$ in the bottom one to remove the effect of different redshifts. Gray dots indicate the whole galaxy sample. Blue dots mark bright objects with $\mathrm{F200W}<28$ mag in the redshift interval $3 \leq z \leq 6.5$ of interest, as labeled. Large filled circles indicate our selected QGs at $z>3$ using the $NUVUVJ$ diagram with their uncertainties, color coded according to the nominal $P_{\rm Q}$ in the run with the full filter set.}
    \label{fig:filters}
\end{figure*}

We tested our sample selection against the availability of \hst\ filters sampling the rest-frame UV/optical emission at $z>3$. As mentioned in Section \ref{subsec:sanity}, we refitted the photometry in CEERS and PRIMER retaining only the available NIRCam wide filters (F090W, F115W, F150W, F200W, F277W, F356W, and F444W). F090W is not available in CEERS, which thus constitutes a more extreme test of the coverage of $NUV$ at $z>3$. In Figure \ref{fig:filters} we show the rest-frame $NUV$, $U$, $V$, and $J$ flux densities in the two fitting runs, also removing the effect of $z_{\rm phot}$. The results are fully consistent -- and more so when F090W is included, as in the case of the test on PRIMER.

\subsection{Sub-millimetric coverage and spectroscopically confirmed objects}
\label{subappendix:submm_spec}
We cross-checked our list of candidate quiescent objects with available catalogs of sub-millimetric surveys in  CEERS ($450$ and $850$ \um\ down to $\sigma_{450}=1.2$ and $\sigma_{850}=0.2$ mJy beam$^{-1}$ with Scuba-2 in the deep tier of the S2CLS survey, \citealt{zavala_2017}; $\sigma_{850}=0.2$ mJy beam$^{-1}$ over the full survey, \citealt{geach_2017}), PRIMER ($870$ \um\ with ALMA from the AS2UDS survey targeting Scuba-2 sub-millimeter galaxies from the S2CLS survey, \citealt{geach_2017}, and detecting sources as faint as $0.6$ mJy at $>4.3\sigma$, \citealt{dudzeviciute_2020}; \citealt{cheng_2022_pearls} based on a combination of archival data), and SMACS0723 ($1.1$ mm observations with $\sigma_{1.1\rm mm}=66.1$ $\mu$Jy beam$^{-1}$ from ALMA in the context of the ALCS Survey, \citealt{kokorev_2022}; S. Fujimoto et al. in preparation). These limits correspond to $\mathrm{SFR}=33-26$ (CEERS/S2CLS-deep), $200-150$ (S2CLS shallow); $25-19$ (PRIMER/AS2UDS), $130-102$ (S2CLS), and $25-16$ \myr\ (SMACS0723/ALCS) at $z=3-6$, obtained by rescaling the $1\sigma$ rms with a modified black body with temperature $T_{\rm dust}=40$~K, $\beta=2$, $k_0=0.43\,\mathrm{cm^2\,g^{-1}}$ at $\lambda_0=850$ $\mu$m \citep{li_2001}, and accounting for the lesser effect of the  CMB \citep{dacunha_2013}. The S2CLS (shallow) survey covers all the CEERS and PRIMER fields. The deeper portion of the survey described in \cite{zavala_2017} covers approximately 45\% of our final samples in CEERS. The ALCS coverage of SMACS0723 is of $\sim3$ arcmin$^2$ centered on the cluster. AS2UDS and the ALMA archival observations are pointed and covered an area $\sim 600\times$ smaller than the parent S2CLS survey in the UDS field ($0.96$ deg$^2$, \citealt{stach_2019}). As mentioned in Section \ref{subsec:sanity}, we retrieve one $\sim5\sigma$-detection at $850$ \um\ from Scuba-2 at a $0\farcs9$ distance from a candidate $UVJ$ quiescent galaxy at $z=3.54$ in CEERS (S2CLS-EGS-850.063 in \citealt{zavala_2017}, \#9329 in our catalog). This candidate is selected by virtue of its uncertainty on the $V-J$ color (0.4 mag) and the introduction of a padded box, while it is not picked by the $NUVUVJ$ criterion. However, several other possible optical/near-IR counterparts fall within the Scuba-2 beam (S. Gillman et al. in preparation), making the physical association inconclusive. Moreover, we matched our candidates with a compilation of spectroscopically confirmed galaxies from the literature. Despite the scarcity of these spectroscopic samples, we retrieve all sources in CEERS in both our selections from \cite{schreiber_2018c} and with fully consistent $z_{\rm phot}$ (($z_{\rm spec}$, $z_{\rm phot}$): EGS-18996: ($3.239$, $3.12^{+0.09}_{-0.05}$); EGS-40032: ($3.219$, $3.35^{+0.09}_{-0.11}$; EGS-31322: ($\sim 3.434$, $3.54^{+0.09}_{-0.10}$)). We do not find any further matches with spectroscopically confirmed objects at any redshifts in our archive of Keck/MOSFIRE observations (G. Brammer et al. in preparation, \citealt{valentino_2022}) nor in the 3D-\textit{HST} survey \citep{skelton_2014, momcheva_2016}.

\subsection{Comparison with \jwst-selected photometric quiescent candidates in the literature}
\label{subappendix:carnall}
\begin{figure*}    \includegraphics[width=\textwidth,height=\textheight,keepaspectratio]{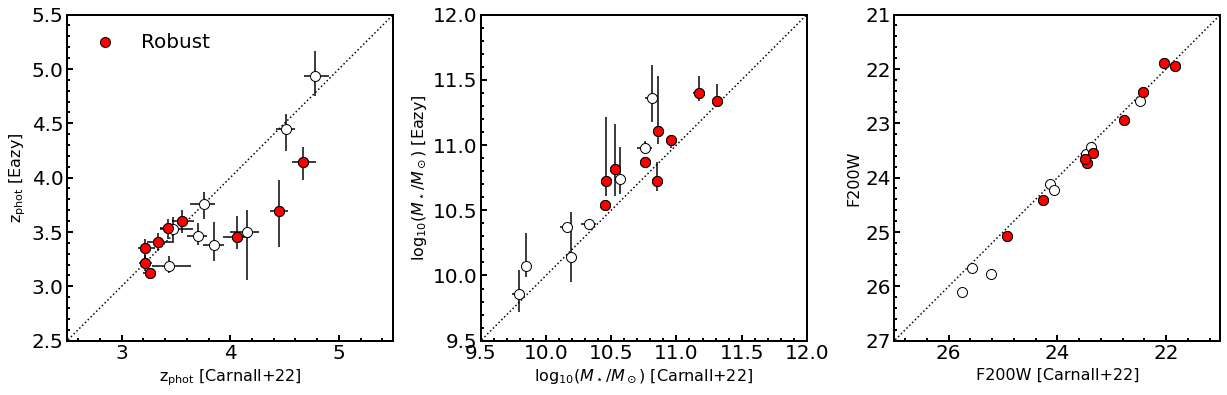}
    \caption{Comparison with \jwst-selected QG at $z>3$ in CEERS. Photometric redshift \zphot, \mstar, and NIRCam F200W in our analysis (Y-axis) and in that presented in \cite{carnall_2022qg} (X-axes). Open and red circles indicate the full and ``robust'' samples of candidates identified by the authors in the CEERS field. We did not apply any correction to homogenize our \cite{chabrier_2003} IMF with that of \cite{kroupa_2001}.}
    \label{fig:carnall}
\end{figure*}

Figure \ref{fig:carnall} shows the comparison between our F200W magnitudes and SED modeling results with \textsc{eazy-py} and those from \cite{carnall_2022qg} for a sample of 17 candidate QGs identified in CEERS by virtue of their low $\mathrm{sSFR}<0.2/t_{\rm obs}$, where $t_{\rm obs}$ is the age of the Universe at the redshift of the galaxy. As mentioned in Section \ref{subsec:sanity}, there is an excellent overlap between our extended $UVJ$ selection and that in \cite{carnall_2022qg}, especially for their ``robust'' sample. Sources \#9844, 4921 in our catalog (78374, 76507 in \citealt{carnall_2022qg}) are excluded by virtue of their blue colors, while \#9131 (92564) has a large uncertainty on $V-J$ ($\sigma_{\rm V-J}=0.62$ mag). The overlap is less extended when imposing $P_{\rm Q,50\%}\geq 0.1$. Sources below this threshold are either at the bluest (\#9844, 4921) or reddest end of the color distribution (e.g., \#7432, 8556 = 40015, 42128), the latter being mainly occupied by dusty SFGs. We remark the fact that our photometry is extracted in $0\farcs5$ apertures and, thus,  traces the properties of the central regions of galaxies. In presence of strong color gradients, as suggested by the RGB images of some of our candidates, photometry in larger apertures or based on surface brightness modeling across bands can drive to different results (e.g., \#7432; see also \citealt{gimenez-arteaga_2022}). Despite this, we find an overall agreement in \zphot\ and \mstar\ (Figure \ref{fig:carnall}). If any, our \zphot\ seem to be systematically lower and \mstar\ larger than those derived by \cite{carnall_2022qg} (\citealt{kocevski_2022} also report lower redshift estimates).
However, these offsets are in the realm of typical statistical and systematic uncertainties that different codes ran with a variety of parameters can produce.\\

In addition, our selections do not retrieve the dusty candidate QG at $z_{\rm phot} \sim 5.4$ in SMACS0723 presented in \citet[ID\#2=KLAMA, \#1536 in our catalog; $\mathrm{R.A.} = 110.70257564$, $\mathrm{Dec.}=-73.48472291$ in the Gaia DR3 astrometric reference]{rodighiero_2023}. Our photometry and SED modeling place this object at $z_{\rm phot}=3.58^{+0.60}_{-0.24}$ and assigns it a $M_\star =3.0^{+1.4}_{-0.8}\times 10^{10}\,M_\odot$ and $P_{\rm Q,50\%}\ll 0.1$.\\

We highlight the fact that the comparison with both these works is partially affected by the different \jwst\ zeropoint photometric calibration, an element in constant evolution to date.

\section{Stellar mass limits}
\label{appendix:mass_completeness}
\begin{figure*}
    \includegraphics[width=\textwidth,height=\textheight,keepaspectratio]{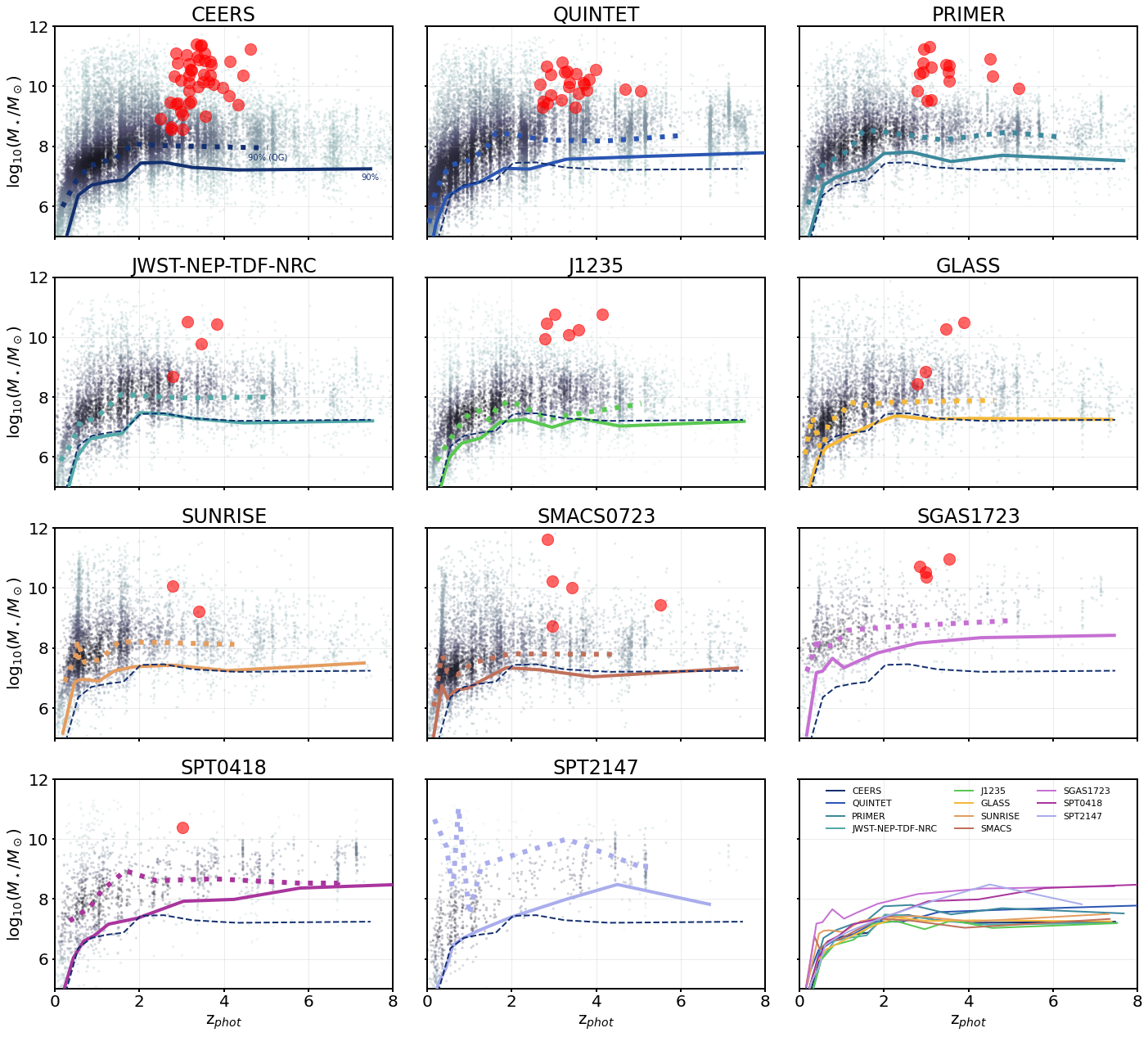}
    \caption{Stellar masses as a function of the photometric redshifts. Gray points indicate sources in each field as label. The color intensity scales as the density of points. The solid color lines mark the 90\% percentile of the \mstar\ distribution in redshift bins of equal number of points (i.e., 90\% of the galaxies lie above these lines in each bin). For reference, the dotted lines indicate the percentile of the \mstar\ distribution for $UVJ$-selected red galaxies in the catalog at any redshifts (i.e., 90\% of the $UVJ$-selected QGs at any redshifts lie above these lines in each bin). We show the 90\% percentile of the distribution of galaxies in the CEERS catalog in each panel (dashed blue line). A direct comparison of the 90\% percentiles is shown in the bottom right panel. For reference, the red circles show the location of our visually-inspected $UVJ$-selected sample of quiescent candidates at $3<z<6.5$.}
    \label{fig:mass_completeness}
\end{figure*}

Different mass limits could be a concern to draw comparison among fields with uneven photometric coverage and depth. In Figure \ref{fig:mass_completeness}, we show that our comparison is robust at the masses considered here. While a full-fledged analysis of the galaxy populations in our catalogs and stellar mass functions is deferred to future work, we show that the stellar mass cuts adopted in Section \ref{sec:number_densities} are well above the threshold where completeness is expected to drop. For reference, we show our visually inspected $UVJ$ sample, more massive that the limit set by the 90\% percentile of the \mstar\ distribution in the $3<z<6.5$ redshift interval in every field. This holds also for the robust $NUVUVJ$-selected sample.\\

\begin{figure}
    \includegraphics[width=\columnwidth]{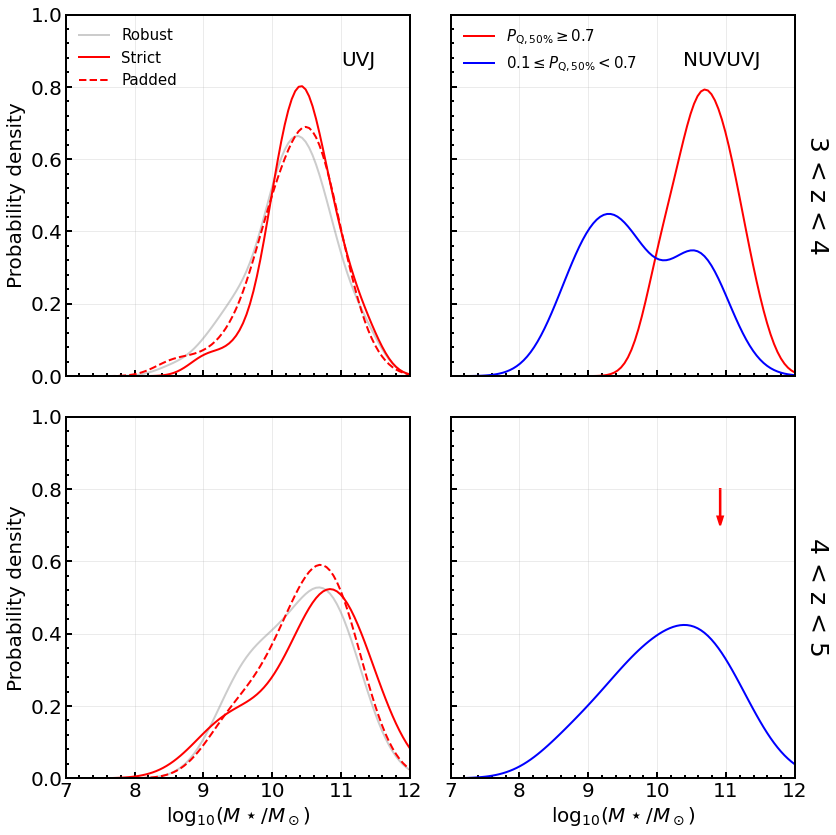}
    \caption{Stellar mass distributions of the selected samples. The solid lines mark the probability density functions of \mstar\ for the $UVJ$ and $NUVUVJ$ final samples colored as labeled. The curves are smoothed with a Gaussian Kernel Density Estimator and normalized to an area of 1. \textit{Top row:} Redshift interval $3<z<4$. \textit{Bottom row:} $4<z<5$. Only \# 185 in PRIMER is selected with $P_{\rm Q,\,50\%}\geq0.7$ at $z>4$. The red arrow marks its stellar mass estimate in the bottom right panel.}
    \label{fig:mass_distribution}
\end{figure}

Figure \ref{fig:mass_distribution} shows the \mstar\ distributions of the $UVJ$ and $NUV-U,\,V-J$ selected sample. As noted in Section \ref{subsec:overlap}, candidates with $P_{\rm Q,\,50\%}\geq0.7$ maximally overlap with traditionally selected $UVJ$ galaxies. Both selection criteria tend to pick the most massive objects. Lower $P_{\rm Q,\,50\%}$ thresholds also select bluer and less massive objects that might have recently quenched. We also note that our ``strict'' and ``padded'' subsamples of the overall robust, but looser visually inspected $UVJ$ pool of candidates do not introduce any immediately evident bias in mass.

\section{Literature compilation}
\label{app:literature}
Table \ref{tab:literature} includes complementary information on the comoving number densities of massive quiescent galaxies at $3\lesssim z \lesssim 4$ reported in Figure \ref{fig:literature}. The estimates are homogenized in terms of redshift interval and above the same mass limit of $\mathrm{log}(M_\star/M_\odot)\gtrsim 10.6$ to the best of our knowledge. A lesser $0.1$ dex difference in the mass integration limit remains for the estimates in \cite{schreiber_2018c} and \cite{cecchi_2019}, while the redshift interval considered in \cite{straatman_2014} is $3.4\leq z < 4.2$. The areas covered by the observations are those quoted in the original papers. There the reader can find further details about the exact selection techniques and their refinements, while we report the primary criterion in the table.

\begin{deluxetable*}{lrrcccc}
    \tablecaption{Comoving number densities of quiescent galaxies at $3<z<4$ in the literature.\label{tab:literature}}
    \tablehead{
    \colhead{Label}&
    \colhead{$n$}&
    \colhead{Area/Box side}&
    \colhead{Primary selection}&
    \colhead{$\mathrm{log}(M_\star)$ limit}&
    \colhead{IMF}&
    \colhead{References}\\ 
    \colhead{}&
    \colhead{\small [Mpc$^{-3}$]}&
    \colhead{}&
    \colhead{}&
    \colhead{\small [$M_\odot$]}&
    \colhead{}&
    \colhead{}
    }
    \startdata
    \multicolumn{7}{c}{Observations}\\
    \tableline
    Muzzin+13&    $2.7\times10^{-6}$&               $1.62$ deg$^2$&    $UVJ$ (SMF)&      $10.6$&    Kroupa&      M13, V20\\
    Straatman+14& $1.8^{+0.7}_{-0.7}\times10^{-5}$& $121$ arcmin$^2$&  $UVJ$&            $10.6$&    Chabrier&    S14\\
    Davidzon+17&  $1.5^{+0.6}_{-0.7}\times10^{-6}$& $0.62$ deg$^2$&    $NUVrJ$ (SMF)&    $10.6$&    Chabrier&    D17, V20\\
    \multirow{2}{*}{Schreiber+18}& $1.4^{+0.3}_{-0.3}\times10^{-5}$&  \multirow{2}{*}{$442$ arcmin$^2$}& $UVJ$& \multirow{2}{*}{$10.5$}& \multirow{2}{*}{Chabrier} & \multirow{2}{*}{S18}\\
    & $2.0^{+0.3}_{-0.3}\times10^{-5}$& & sSFR&  &    & \\
    \multirow{2}{*}{Merlin+19}& $4.9^{+0.6}_{-0.6}\times10^{-6}$& \multirow{2}{*}{$970$ arcmin$^2$}& SED  (w/lines)&    \multirow{2}{*}{$10.6$}& \multirow{2}{*}{Chabrier} & \multirow{2}{*}{M18, M19}\\
    & $2.0^{+0.1}_{-0.1}\times10^{-5}$& & SED  (complete)&  &    & \\
    Cecchi+19& $1.0^{+0.3}_{-0.3}\times10^{-6}$& $1.38$ deg$^2$& $NUVrJ$+sSFR&    $10.5$&    Chabrier& C19, This work\\    
    Girelli+19& $1.5^{+0.5}_{-0.7}\times10^{-6}$& $1.38$ deg$^2$& Observed colors&    $10.6$&    Chabrier& G19, V20\\   
    Shahidi+20& $4.9^{+1.6}_{-1.2}\times10^{-6}$& $964$ arcmin$^2$& Balmer, $UVJ$, SED&    $10.6$&    Chabrier& S20, This work\\    
    \multirow{2}{*}{Carnall+20}& $1.1^{+0.4}_{-0.3}\times10^{-5}$& \multirow{2}{*}{$370$ arcmin$^2$}& sSFR (Full)&    \multirow{2}{*}{$10.6$}& \multirow{2}{*}{Kroupa} & \multirow{2}{*}{C20, This work}\\
    & $5.1^{+3.0}_{-2.0}\times10^{-6}$& & sSFR (Robust)&  &    & \\
    Weaver+22& $9.4^{+1.7}_{-1.7}\times10^{-6}$& $1.27$ deg$^2$& $NUVrJ$&    $10.6$&    Chabrier& W22, This work\\    
    Gould+23& $1.2^{+0.4}_{-0.4}\times10^{-5}$& $1.27$ deg$^2$& GMM&    $10.6$&    Chabrier& G23\\    
    \multirow{2}{*}{Carnall+22}& $6.3^{+3.8}_{-2.5}\times10^{-5}$& \multirow{2}{*}{$30$ arcmin$^2$}& sSFR (Full)&    \multirow{2}{*}{$10.6$}& \multirow{2}{*}{Kroupa} & \multirow{2}{*}{C22, This work}\\
    & $4.2^{+3.3}_{-2.0}\times10^{-5}$& & sSFR (Robust)&  &    & \\    \tableline
    \tableline
    \multicolumn{7}{c}{Simulations}\\
    \tableline
    \multirow{2}{*}{Illustris-1} & $3.3^{+0.8}_{-0.8}\times10^{-6}$&  \multirow{2}{*}{$107$ cMpc}& sSFR ($z=3.0$)& \multirow{2}{*}{$10.6$}&    \multirow{2}{*}{Chabrier}& \multirow{2}{*}{V20}\\
      & $<8.1\times10^{-7}$& & sSFR ($z=3.7$)& &    & \\
    \multirow{2}{*}{Illustris TNG-100} & $5.4^{+0.0}_{-1.0}\times10^{-5}$& \multirow{2}{*}{$107$ cMpc}& sSFR ($z=3.0$)& \multirow{2}{*}{$10.6$}&    \multirow{2}{*}{Chabrier}& \multirow{2}{*}{V20}\\
     & $7.8\times10^{-6}$& & sSFR ($z=3.7$)& &   & \\
    \multirow{2}{*}{Illustris TNG-300}& $3.0^{+0.0}_{-0.5}\times10^{-5}$& \multirow{2}{*}{$293$ cMpc}& sSFR ($z=3.0$)& \multirow{2}{*}{$10.6$}&    \multirow{2}{*}{Chabrier}& \multirow{2}{*}{V20}\\
    & $2.6^{+0.0}_{-0.5}\times10^{-6}$& & sSFR ($z=3.7$)& & & \\  
    \multirow{2}{*}{EAGLE}&  $1.0\times10^{-6}$&  \multirow{2}{*}{$100$ cMpc}& sSFR ($z=3.0$)& \multirow{2}{*}{$10.6$}&    \multirow{2}{*}{Chabrier}& \multirow{2}{*}{This work}\\
    &  $<1.8\times10^{-6}$& & sSFR ($z=3.9$)& &    & \\    
    \enddata
    \tablecomments{\textbf{Primary selection:} $UVJ$, $NUVrJ$ = rest-frame color diagrams; SMF = integration of stellar mass functions (from best-fit parameters); sSFR = cuts in specific star formation rates estimated via SED modeling; SED = composite cuts based on SED modeling; observed colors; Balmer = Balmer break; GMM = Gaussian Mixture Modeling in the $NUV-U$, $U-V$, $V-J$ space.
    \textbf{Initial Mass Function:} \cite{kroupa_2001}, \cite{chabrier_2003}. We do not apply any mass correction to convert one IMF to the other.
    \textbf{References:} M13 = \cite{muzzin_2013s}; S14 = \cite{straatman_2014}; D17 = \cite{davidzon_2017}; S18 = \cite{schreiber_2018c}; M18 = \cite{merlin_2018}; M19 = \cite{merlin_2019}; C19 = \cite{cecchi_2019}; G19 = \cite{girelli_2019}; V20 = \cite{valentino_2020a}; S20 = \cite{shahidi_2020}; C20 = \cite{carnall_2020}; W22b = \cite{weaver_2022_smf}; G23 = \cite{gould_2022}; C22 = \cite{carnall_2022}.}
\end{deluxetable*}

\begin{deluxetable*}{lccccc}
    \tablecaption{Comoving number densities of quiescent galaxies at $3<z<4$.\label{tab:numberdensities_individual_field}}
    \tablehead{
    \colhead{Field}&
    \colhead{Area}&
    \colhead{$UVJ$}&
    \colhead{$UVJ$}&
    \colhead{$NUVUVJ$}&
    \colhead{$\sigma_{\rm CV}$}\\ 
    \colhead{}&
    \colhead{\small [arcmin$^{2}$]}&
    \colhead{\small Strict}&
    \colhead{\small Padded}&
    \colhead{\small $P_{\rm Q,50\%}$}&
    \colhead{$\%$}
    }
    \startdata
    \multirow{2}{*}{CEERS} & \multirow{2}{*}{$34.7$}& $3.2^{+2.8}_{-1.6}$& $4.0^{+3.0}_{-1.8}$& $3.9^{+2.9}_{-1.8}$&  0.29 \\
    & & $6.6^{+3.5}_{-2.4}$& $7.5^{+3.6}_{-2.6}$& $7.6^{+3.6}_{-2.6}$& 0.53 \\  
    \multirow{2}{*}{Stephan's Quintet} & \multirow{2}{*}{$35.0$}& $6.4^{+3.4}_{-2.4}$& $7.4^{+3.6}_{-2.5}$& $3.9^{+2.9}_{-1.8}$&  0.30 \\
    & & $0.0^{+1.7}_{-0.0}$& $0.0^{+1.8}_{-0.0}$& $0.0^{+1.7}_{-0.0}$& 0.55 \\    
    \multirow{2}{*}{PRIMER} & \multirow{2}{*}{$21.9$}& $2.9^{+3.8}_{-1.9}$& $3.0^{+3.8}_{-1.9}$& $1.6^{+3.4}_{-1.3}$&  0.32 \\
    & & $3.4^{+4.0}_{-2.1}$& $4.5^{+4.3}_{-2.4}$& $3.4^{+4.0}_{-2.1}$& 0.58\\    
    \multirow{2}{*}{NEP} & \multirow{2}{*}{$9.7$}& $7.5^{+8.9}_{-4.6}$& $7.5^{+8.9}_{-4.6}$& $5.0^{+8.1}_{-3.6}$&  0.34 \\
    & & $0.0^{+6.0}_{0.0}$& $0.0^{+6.6}_{0.0}$& $0.0^{+6.0}_{0.0}$& 0.61\\    
    \multirow{2}{*}{J1235} & \multirow{2}{*}{$9.0$}& $0.3^{+6.7}_{-0.3}$& $0.3^{+6.7}_{-0.3}$& $3.7^{+8.2}_{-3.0}$&  0.34 \\
    & & $0.0^{+6.5}_{-0.0}$& $0.3^{+6.6}_{-0.3}$& $0.3^{+6.6}_{-0.3}$& 0.62 \\    
    \multirow{2}{*}{GLASS} & \multirow{2}{*}{$8.5$}& $0.0^{+6.9}_{-0.0}$& $0.0^{+6.9}_{-0.0}$& $0.0^{+6.9}_{-0.0}$&  0.34 \\
    & & $0.0^{+6.9}_{-0.0}$& $0.0^{+6.9}_{-0.0}$& $0.0^{+6.9}_{-0.0}$& 0.62 \\    
    \multirow{2}{*}{Sunrise} & \multirow{2}{*}{$7.3$}& $0.0^{+8.1}_{-0.0}$& $0.0^{+8.1}_{-0.0}$& $0.0^{+8.1}_{-0.0}$&  0.34 \\
    & & $0.0^{+8.1}_{-0.0}$& $0.0^{+8.1}_{-0.0}$& $0.0^{+8.1}_{-0.0}$&0.62 \\    
    \multirow{2}{*}{SMACS0723} & \multirow{2}{*}{$6.5$}& $6.2^{+11.7}_{-4.7}$& $6.2^{+11.7}_{-4.7}$& $4.0^{+10.9}_{-3.5}$&  0.34 \\
    & & $0.8^{+9.5}_{-0.8}$& $0.8^{+9.5}_{-0.8}$& $0.0^{+9.0}_{-0.0}$& 0.63\\    
    \multirow{2}{*}{SGAS1723} & \multirow{2}{*}{$5.3$}& $5.4^{+13.5}_{-4.6}$& $1.9^{+12.0}_{-1.9}$& $0.0^{+11.0}_{-0.0}$&  0.35 \\
    & & $6.1^{+13.8}_{-5.0}$& $6.1^{+13.8}_{-5.0}$& $0.0^{+11.0}_{-0.0}$& 0.64\\   
    \multirow{2}{*}{SPT0418} & \multirow{2}{*}{$5.0$}& $2.5^{+13.0}_{-2.5}$& $2.5^{+13.0}_{-2.5}$& $0.0^{+11.7}_{-0.0}$&  0.35 \\
    & & $0.0^{+11.7}_{-0.0}$& $0.0^{+11.7}_{-0.0}$& $0.0^{+11.7}_{-0.0}$& 0.65\\   
    \multirow{2}{*}{SPT2147} & \multirow{2}{*}{$2.3$}& $0.0^{+25.4}_{-0.0}$& $0.0^{+25.4}_{-0.0}$& $0.0^{+25.4}_{-0.0}$&  0.37 \\
    & & $0.0^{+25.4}_{-0.0}$& $0.0^{+25.4}_{-0.0}$& $0.0^{+25.4}_{-0.0}$& 0.67 \\  
    \tableline
    \multirow{2}{*}{Combined} & \multirow{2}{*}{$145.1$}& $3.9^{+1.2}_{-0.9}$& $4.1^{+1.2}_{-0.9}$& $2.8^{+1.0}_{-0.8}$& 0.10  \\
    & & $2.4^{+1.0}_{-0.7}$& $2.7^{+1.0}_{-0.8}$& $2.3^{+1.0}_{-0.7}$& 0.18\\
    \enddata
    \tablecomments{The comoving number densities are expressed in units of $10^{-5}$ Mpc$^{-3}$. Each field has two entries: the first and second rows refer to the $\mathrm{log}(M_\star/M_\odot)=[9.5,10.6)$ and the $\geq10.6$ bins, respectively. The uncertainties reflect the Poissonian $1\sigma$ confidence interval. Upper limits are at $1\sigma$ using the same approach \citep{gehrels_1986}. Statistical uncertainties are accounted by integrating the $p(z)$ within $3<z<4$. The uncertainties due to cosmic variance are expressed as fractional $\sigma_{\rm CV}$ deviations (Section \ref{subsec:cv}). The selections are described in Section \ref{sec:sampleselection}. The adopted threshold for the $NUVUVJ$ selection is $P_{\rm Q,50\%}\geq 0.1$.}
\end{deluxetable*}

\bibliography{bib_atlas_qg}
\bibliographystyle{aasjournal}

\end{document}